\documentclass[aps,pre,twocolumn,amsmath,amssymb,showpacs,amsfonts]{revtex4}
\usepackage{graphicx}
\usepackage{dcolumn}
\usepackage{bm}
\usepackage{color}

\begin{document}

\title{Convective stability of turbulent Boussinesq flow in the dissipative range and flow around small particles}

\author{Itzhak Fouxon}
\author{Alexander Leshansky}\email{lisha@technion.ac.il}

\affiliation{Department of Chemical Engineering 
Technion -- IIT, Haifa 32000, Israel}

\begin{abstract}

We consider arbitrary, possibly turbulent, Boussinesq flow which is smooth below a dissipative scale $l_d$. It is demonstrated that the stability of the flow with respect to growth of fluctuations with scale smaller than $l_d$ leads to a non-trivial constraint. That involves the dimensionless strength of fluctuations of the gradients of the scalar in the direction of gravity $Fl$ and the Rayleigh scale $L$ depending on the Rayleigh number $\mathrm{Ra}$, the Nusselt number $\mathrm{Nu}$ and $l_d$. The constraint implies that the stratified fluid at rest, which is linearly stable, develops instability in the limit of large $\mathrm{Ra}$. This limits observability of solution for the flow around small swimmer in quiescent stratified fluid that has closed streamlines at scale $L$ [A. M. Ardekani and R. Stocker, Phys. Rev. Lett. {\bf 105}, 084502 (2010)]. Correspondingly to study the flow at scale $L$ one has to take turbulence into account. We demonstrate that the resulting turbulent flow around small particles or swimmers can be described by scalar integro-differential advection-diffusion equation. Describing the solutions we show that closed streamlines persist with finite probability. Our results seem to be the necessary basis in understanding flows around small particles and swimmers in natural marine environments.

\end{abstract}

\maketitle

\section{Introduction}

Turbulent flows are considered to be smooth in the dissipation range of scales \cite{Frisch}. Below the viscous scale of the Navier-Stokes turbulence the internal friction of the fluid regularizes the irregular turbulent fluctuations in the inertial range turning the flow into laminar and differentiable. In the case of the Bousinesq turbulence the regularization is done both by viscosity and heat conduction. We define $l_d$ as the dissipative scale below which both the flow and the scalar field coupled to it are smooth. In both cases the self-consistency of laminarity below $l_d$ demands decay of perturbations at scales smaller than $l_d$.

In the Navier-Stokes turbulence the value of the viscous scale $l_d$ can be obtained from the demand that the viscous dissipation balances the non-linear advection term in the equation at this scale. Otherwise said $l_d$ is fixed by demanding that the Reynolds number based on this scale is of order one \cite{Frisch}. That guarantees that the flow below $l_d$ is laminar smooth flow. In particular if one studies the temporal development of wave-packets with small amplitude sized below $l_d$ then the smallness of the Reynolds number based on the packet's size guarantees that the fluctuation decays. Thus the assumption of the flow laminarity below the scale $l_d$ is self-consistent.

Consider now the Bousinesq turbulence that involves interaction of the flow with the scalar. Using the velocity equation one can readily determine the scale $l_{\nu}$ at which the viscous dissipation balances the non-linear advection term in the equation. Similarly, using the scalar equation one can readily determine the scale $l_{dif}$ at which the heat conduction balances the non-linear advection term in the equation. However, in contrast to the Navier-Stokes turbulence, this does not guarantee that the wave-packets with size smaller than $l_d=\min[l_{\nu}, l_{dif}]$ would decay. This is due to the coupling of the flow to the scalar field that produces the mechanism for the growth of fluctuations which is the same mechanism that underlies the convective instability. Thus the demand that the heat conduction and viscosity terms dominate the advection ones in velocity and scalar equations separately does not guarantee the stability yet.

In this work we study the equations governing small perturbations at scales smaller than $l_d$ and demonstrate that the condition of decay of small fluctuations in the dissipation range produces a non-trivial inequality. This inequality can be considered as the condition of convective stability in the dissipative range.

It is a remarkable consequence of the derived condition that the familiar linearly stable state of the vertically stratified fluid at rest turns unstable when Rayleigh number, $\mathrm{Ra}$, measuring the relative importance of buoyancy and dissipative processes in the fluid, is large. This is based on the constraint's consequence that unless the scalar gradient points strictly in the direction of the gravity, the flow gets unstable in the limit $\mathrm{Ra}\to\infty$. Since in natural situations
the direction of the scalar gradient fluctuates then the instability holds. The potential far-reaching consequences of this instability are to be studied.

Further we demonstrate that in the case of the turbulent Boussinesq flow, the condition is obeyed by the known phenomenological relations. In the case of non-small Prandtl number the flow parameters fit the condition precisely. It follows that the values of the dissipative scales can be considered as the condition of stability like in the Navier-Stokes turbulence. In contrast to that, in the case of small Prandtl numbers there is a gap between the dissipative scales prescribed by the stability and those observed phenomenologically. The understanding of this gap demands further study.

Finally we use the stability condition to study the Boussinesq flow around small particles. Although the problem of translation of small particles through the viscous fluid has been studied intensively for centuries already \cite{HB}, unexpectedly, a new result was found recently. Non-trivial flow pattern was predicted for passive particles and self-propelled swimmers that move in the quiescent fluid when fluid's density stratification (present invariably in many aquatic environments) is taken into account \cite{Ardekani}. Stratification produces intrinsic scale $L=(\nu\kappa/\gamma g)^{1/4}$ that characterizes interaction of dissipative processes (characterized by kinematic viscosity $\nu$ and heat conductivity $\kappa$) and gravity $g$, with $\gamma=(1/\rho_0) (-d\rho/dz)$ being the density gradient normalized by the reference density. In the case of stable convection in the horizontal layer of a fluid of depth $d$, $L^4=d^4/\mathrm{Ra}$. Thus below we refer to $L$ as the Rayleigh scale. In particular, the well-known threshold for convective instability corresponds to Rayleigh scale becoming smaller than $d$ times a constant of order one depending on the type of boundary conditions \cite{Drazin}.

In the fluid at rest the Rayleigh scale is the scale at which the impact of stratification on the flow around small particle becomes appreciable. At smaller scales $a\leq r\ll L$ where $a$ is the particle's radius, the stratification is negligible and the usual Stokes flow holds. However, at scales of order $L$ and larger instead of the Stokes's streamlines that are open to infinity one finds closed streamlines and toroidal eddies \cite{Ardekani}. We show here that the flow decay at $r\gtrsim L$ is exponential-like. This is in sharp contrast to the slow algebraic (inverse with the distance) decay of the Stokes flow.

The obtained flow demonstrates that in spite of wide separation of scales of stratification (kilometers) and of particles or self-propelled microorganisms ($0.1-1$~mm) in typical marine environment,
the stratification cannot be neglected - the scale $L$ is of the order of $\sim 1$~mm. It was suggested that this previously unnoticed feature of the flow around small particles in the ocean
may affect propulsion of small organisms and sinking of marine snow particles, diminish the effectiveness of mechanosensing in the ocean \cite{Ardekani}, stifle nutrient uptake of small motile organisms \cite{dsa12} or potentially hinder the drift-induced biogenic mixing \cite{biomix}.

The derivation of the above stratified flow in \cite{Ardekani} presumed constant gradient of the stratified agent. Correspondingly the obtained flow is observable if $L\ll l_{dif}$ where $l_{dif}$ is the scale of spatial variations of the gradients. However, the stability condition derived in this work shows that quiescent stratified fluid with $L\ll l_{dif}$ is stable only if the deviations $(\gamma-\gamma_z)/\gamma$ of the scalar gradient $\bm \gamma$ from the direction of gravity, $\gamma=\gamma_z$, are bounded by small parameter $(L/l_{dif})^4$. Since for  $L/l_{dif}\ll 1$ the parameter $(L/l_{dif})^4$ is vanishingly small then it seems that the predicted flow is unlikely to occur in nature.

To resume, the problem with observability of the solution in \cite{Ardekani}, is that the flow involves intrinsic scale of the fluid $L$ that is independent of the particles' size (in contrast to, say, the Oseen scale). Then the condition of observability of the flow round the particle is the condition on the fluid flow itself, which realization is questionable due to convective instability holding when the gradients of the scalar have natural fluctuations from the vertical direction. We point out that it is not that the particle drives the rest state of the fluid considered in \cite{Ardekani} unstable, but rather that the rest state with considered parameters is not likely to occur in marine environments to begin with.

The stability condition implies that the consideration of the flow round small particles and swimmers in the quiescent stratified fluid has natural applications only in the case $L\gtrsim l_{dif}$ (unless there is strong limitation on the fluctuation of the gradients described in the previous paragraph). However, in order to consider the flow at scale $L$ in this case, one has to take into the spatial variation of the gradient of the background density. This complication both hinders analytic progress and makes the study non-universal. The flow would strongly depend on the way the gradient varies.

Thus we consider the universal situation where the flow is turbulent, which is by far the most common case in nature. In this case the stability condition implies that $L\gtrsim l_{dif}$. We consider the practically important case of large Prandtl numbers where the known phenomenological relations imply $L\sim l_{dif}$. We demonstrate that in this case the calculation of the flow is reduced to the study of scalar advection-diffusion equation. This equation is integro-differential.

Despite the presence of turbulence, to leading order the flow at $a\leq r\ll L$ is still the Stokes flow. We use the advection-diffusion equation to find the correction to that flow due to turbulence. We succeed in finding exact formula for the magnitude of fluctuations due to turbulence in terms of the energy dissipation rate $\epsilon$.

Finally we construct the asymptotic solutions to the flow at the scale $L$. These demonstrate that with finite probability there are closed streamlines.

It should be emphasized that the results are obtained without modelling the statistics of turbulence and they can be directly applied to natural environments.

\section{Behavior of infinitesimal small-scale perturbations in the Boussinesq flow}

In this Section we consider the behavior of infinitesimal small-scale perturbations in the Boussinesq flow that describes interaction of the flow with the stratifying agent, by \cite{Landau,Drazin}
\begin{eqnarray}&&
\partial_t\bm v+\bm v\cdot\nabla\bm v=-\nabla p_0+\phi \bm g+\nu\nabla^2\bm v+\bm f,\ \ \nabla\cdot\bm v=0,\nonumber\\&&
\partial_t\phi+\bm v\cdot\nabla\phi=\kappa\nabla^2\phi, \label{basic0}
\end{eqnarray}
where $\bm v$ is the fluid velocity, $\phi$ the fluid density normalized by the reference value $\rho_0$, $p_0$ is the properly defined pressure \cite{Drazin} and $\bm g=-g{\hat z}$ is the gravitational acceleration.
The force $\bm f$ represents either a body force and/or a boundary force. We demonstrate that there is a range of parameters for which the infinitesimal small-scale fluctuations would grow exponentially. It is to be stressed that besides smoothness (existence of finite scales of temporal and spatial variation of the fields) no assumptions are made on the flow. In particular, the flow can be turbulent, which is the typical case in natural marine environments.

We briefly review the characteristic properties of the turbulent flow governed by Eqs.~(\ref{basic0}). The smallest scale of fluctuations of $\phi$ is the diffusive scale $l_{dif}$. Though $\phi$ itself is dominated by the large-scale fluctuations at the outer scale of the flow $L_0$, the fluctuations of the gradients $\nabla\phi$ are dominated by the scale $l_{dif}$. The latter typically is much smaller than $L_0$. This is similar to properties of velocity in the Navier-Stokes turbulence \cite{Frisch} where the velocity is dominated by $L_0$ but its gradients by $l_{\nu}\ll L_0$. Here $l_{\nu}\sim \nu^{3/4}\epsilon^{-1/4}$  where $\epsilon$ is the rate of viscous energy dissipation per unit volume of the fluid see \cite{Frisch}). In fact within the Kolmogorov theory the scalings of velocity and the scalar coincide in the inertial range, see \cite{Lohse} and references therein. The scale $l_{dif}$ is determined by balancing the advection and diffusion terms in Eqs.~(\ref{basic0}). The balance gives that $\delta v(l_{dif})l_{dif}\sim \kappa$ where $\delta v(l_{dif})$ is the typical difference of velocities at spatial scale $\sim l_{dif}$. Similarly the smallest spatial scale of variations of $\bm v$, the viscous scale $l_{\nu}$ is determined from $\delta v(l_{\nu})l_{\nu}\sim \nu$.
If the Prandtl number $\mathrm{Pr}=\nu/\kappa$ obeys $\mathrm{Pr}\gtrsim 1$ then the viscous scale $l_{\nu}$, below which the flow is smooth, obeys $l_{\nu}\gtrsim l_{dif}$, so that at the scale $l_{dif}$ one can approximate $\delta v(l_{dif})$ by its derivative, $\delta v(l_{dif})\sim \lambda l_{dif}$ where $\lambda$ is the typical value of the velocity gradient. The latter is given by $\lambda=\sqrt{\epsilon/\nu}$. It follows that $l_{dif}=\sqrt{\kappa/\lambda}$ if $\mathrm{Pr}\gtrsim 1$. In the case $\mathrm{Pr}\ll 1$ the scale $l_{dif}$ belongs to the inertial range of turbulence producing the Kolmogorov theory estimate $l_{dif}\sim \kappa^{3/4}\epsilon^{-1/4}=\mathrm{Pr}^{-3/4}l_{\nu}$ where $\delta v(l_{dif})\sim \epsilon^{1/3}l_{dif}^{1/3}$. Similar estimates can be written for Bolgiano-Obukhov scaling.

Independently of $\mathrm{Pr}$ and whether the flow is described by Kolmogorov or Bolgiano-Obukhov scalings, what is relevant to the present study is that $\nabla\phi$ can be considered smooth below scales $l_{dif}$ so that it is approximately constant over distances much smaller than $l_{dif}$. The temporal scale $t_{dif}$ of variations of $\nabla\phi$  is $\lambda^{-1}$ if $\mathrm{Pr}\gtrsim 1$ and $[\kappa/\epsilon]^{1/2}=\lambda^{-1} \mathrm{Pr}^{-1/2}$ otherwise (this follows from the estimates above). One can write $t_{dif}^{-1}=\lambda \min[1, \mathrm{Pr}^{1/2}]$.

The time-development of small perturbations is described by linearized version of Eqs.~(\ref{basic0}),
\begin{eqnarray}&&
\partial_t\bm u+\bm v\cdot\nabla\bm u+\bm u\cdot\nabla \bm v=-\nabla p+\theta \bm g+\nu\nabla^2\bm u,\nonumber\\&&
\partial_t\theta+\bm v\cdot\nabla\theta+\bm u\cdot\nabla\phi=\kappa\nabla^2\theta,\ \  \nabla\cdot\bm u=0,\label{o3}
\end{eqnarray}
where $\bm u$, $\theta$ and $p$ are perturbations of velocity, scalar and pressure respectively. We consider the time-development of wave-packets with characteristic wave-length much smaller than the smallest spatial scale $l_d$ of the flow which is the minimum of the diffusive $l_{dif}$ and viscous $l_{\nu}$ scales, $l_d=min[l_{dif}, l_{\nu}]$. Introducing the fields in the frame that moves with wave packets' center $\bm q_{cm}(t)$, that is considering $\bm u'(\bm x, t)= \bm u(\bm x+\bm q_{cm}(t), t)$ and $\theta'(\bm x, t)=\theta(\bm x+\bm q_{cm}(t), t)$ we find the following equation for $\bm u'(\bm x, t)$
\begin{eqnarray}&&\!\!\!\!
\partial_t\bm u'+\left[\bm v(\bm x+\bm q_{cm}(t), t)-\bm v(\bm q_{cm}(t), t)\right]\cdot\nabla\bm u'
\nonumber\\&&\!\!\!\!
+\bm u'\cdot\nabla \bm v(\bm x+\bm q_{cm}(t), t)=-\nabla p'+\theta' \bm g
\nonumber\\&&\!\!\!\!
+\nu\nabla^2\bm u',\ \ \nabla\cdot\bm u'=0,
\end{eqnarray}
where $p'(\bm x, t)=p(\bm x+\bm q_{cm}(t), t)$ and $\partial_t \bm q_{cm}(t)=\bm v(\bm q_{cm}(t), t)$.
The equation for $\theta'(\bm x, t)$ reads
\begin{eqnarray}&&
\partial_t\theta'+\left[\bm v(\bm x+\bm q_{cm}(t), t)-\bm v(\bm q_{cm}(t), t)\right]\cdot\nabla\theta'
\nonumber\\&&
+\bm u'\cdot\nabla\phi(\bm x+\bm q_{cm}(t), t)=\kappa\nabla^2\theta'.
\end{eqnarray}
We now use that the wave packet's size is much smaller than $l_d$ so that one can use the Taylor series which to leading order gives
\begin{eqnarray}&&
\partial_t\bm u'+\sigma \bm x\cdot\nabla\bm u'+\sigma \bm u'=-\nabla p'+\theta' \bm g+\nu\nabla^2\bm u',\nonumber\\&&
\partial_t\theta'+\sigma \bm x\cdot\nabla\theta'=\bm u'\cdot\bm \gamma+\kappa\nabla^2\theta',
\end{eqnarray}
where we introduced the matrix of velocity derivatives in the fluid particle's frame, $\sigma_{ik}(t)=\nabla_k v_i(\bm q_{cm}(t), t)$ and the gradient of the scalar in that frame
$\bm \gamma(t)=-\nabla\phi(\bm q_{cm}(t), t)$. The terms $\sigma\bm x$ describe the distortion of the wave-packet due to local velocity gradients.

Finally, we observe that the small size of the wave packet in comparison with $l_d$ implies that the stretching and advection
terms can be neglected in comparison with viscous and heat conduction terms - the wave-packet is in the range dominated by the dissipative viscosity and heat conduction processes. We find
\begin{eqnarray}&&
\partial_t\bm u'=-\nabla p'+\theta' \bm g+\nu\nabla^2\bm u',\ \ \nonumber\\&&
\partial_t\theta'=\bm u'\cdot\bm \gamma+\kappa\nabla^2\theta'. \label{sist}
\end{eqnarray}
The time development of Fourier modes obeys
\begin{eqnarray}&&
\partial_t\bm u'=-i\bm k p' -g \theta' {\hat z}-\nu k^2\bm u',\ \ \bm k\cdot\bm u'=0, \nonumber
\\&&
\partial_t\theta'=\bm u'\cdot\bm \gamma -\kappa k^2\theta'.
\end{eqnarray}
Multiplying the first equation with $\bm k$ we find $ik^2 p'=-g \theta' k_z$ or $p'=i g \theta' k_z/k^2$. Thus
\begin{eqnarray}&&
\partial_t u'_i=g\theta'  k_i k_z/k^2-g \theta' \delta_{iz}-\nu k^2 u'_i,
\end{eqnarray}
which gives
\begin{eqnarray}&&\!\!\!\!\!\!\!\!\!\!\!\!
\left[\partial_t+\nu k^2\right]\left[\partial_t+\kappa k^2\right] u'_i=g\left[ k_i k_z/k^2- \delta_{iz}\right]\bm u'\cdot\bm \gamma.\label{o1}
\end{eqnarray}
This equation by itself cannot be solved due to the time-dependence of $\bm \gamma$ with characteristic time-scale $t_{dif}$ (the problem is similar to one-dimensional Schr\"{o}dinger equation on vector wave-function $\bm u(t)$ with $\gamma(t)$ playing the role of the potential. It can only be solved for particular time-dependencies of $\bm \gamma(t)$.) However, the equation can be solved in the limit where the scale of temporal variations $t_c$ of $u'_i$ obeys $t_c\ll t_{dif}$ where one can use ``adiabatic approximation". To find the range of parameters of this limit, we assume $t_c\ll t_{dif}$ and demand self-consistency. We consider $\bm \gamma$ as slowly varying quantity that can be treated as a constant to the leading order. For constant $\bm \gamma$ taking the scalar product of the above equation with $\bm \gamma$ produces closed equation on $w=\bm v\cdot\bm \gamma$,
\begin{eqnarray}&&
\left[\partial_t+\nu k^2\right]\left[\partial_t+\kappa k^2\right] w=-g {\tilde \gamma}w,\label{o2}
\end{eqnarray}
where ${\tilde \gamma}(\bm k)=\gamma_z-(\bm \gamma \cdot {\hat k}){\hat k}_z$ with ${\hat k}=\bm k/k$. This is the Fourier space version of the corresponding equations for the study of convective instability in \cite{Drazin}. The solutions are proportional to $\exp[\lambda t]$ where
\begin{eqnarray}&&
(\lambda+\nu k^2)(\lambda+\kappa k^2)=-g {\tilde \gamma},
\end{eqnarray}
with the larger solution (if the square root is real, otherwise $Re\lambda<0$)
\begin{eqnarray}&&
\lambda(\bm k)=\frac{1}{2}\left[\sqrt{(\nu-\kappa)^2k^4-4g {\tilde \gamma}}-(\nu+\kappa)k^2\right]. \label{labd}
\end{eqnarray}
Consider the dependence of $\lambda(\bm k)$ on $k$ at fixed orientation ${\hat k}$ so that ${\tilde \gamma}$ is fixed too. In the convection-like case one has ${\tilde \gamma}<0$ (to see that in the case of convection ${\tilde \gamma}<0$, note that $\bm \gamma=-\gamma {\hat z}$, so that ${\tilde \gamma}=-\gamma k_{\perp}^2/k^2$ with $k_{\perp}^2=k^2-k_z^2$) so that the solution is positive at small $k<k_0$ and negative at $k>k_0$, where $\lambda(k=k_0)=0$. We have $k_0^4=-g{\tilde \gamma}/\nu\kappa$. In contrast, in the stratification-like case, ${\tilde \gamma}>0$, the real part of $\lambda$ is always negative (it has unique zero at $k=0$ which has no consequences in physical situations with finite $k$).

It follows from $k_0^4=-g{\tilde \gamma}/\nu\kappa$ that the maximal value of $k$ with positive $\lambda$ is determined by the minimum of ${\tilde \gamma}$ over all possible orientations of ${\hat k}$.
We notice that the maximum of $(\bm \gamma \cdot {\hat k}){\hat k}_z$ over all possible orientations of ${\hat k}$ is given by $(\gamma_z+\gamma)/2$ while the minimum by $(\gamma_z-\gamma)/2$. To see this one can write $\bm \gamma=\gamma_z{\hat z}+\left(\bm \gamma-\gamma_z{\hat z}\right)$ so that
${\hat k_z}\left(\bm \gamma\cdot{\hat k}\right)=\left(\gamma_z+\gamma_{\perp}\tan\delta\right)/(1+\tan^2\delta)$ where $\gamma_{\perp}=\sqrt{\gamma^2-\gamma_z^2}$ and $\cos \delta={\hat k_z}$. Then the extrema over $\delta$ taken at $\tan \delta=\left(-\gamma_z\pm\gamma\right)/\gamma_{\perp}$
are given by the provided expressions. It follows that the minimum of ${\tilde \gamma}=\gamma_z-{\hat k_z}(\bm \gamma\cdot{\hat k})$ is given by
$\gamma_z-(\gamma_z+\gamma)/2=(\gamma_z-\gamma)/2\leq 0$ where the equality holds only if $\bm \gamma$ points precisely in $z-$direction. Thus $\lambda(\bm k)$ is positive for certain modes provided the wave-numbers with
$k^4<g(\gamma-\gamma_z)/2\nu\kappa$ are admissible.

In the case of stratification with $\bm \gamma=\gamma{\hat z}$ where $g(\gamma-\gamma_z)/2\nu\kappa=0$ there are no modes with positive $\lambda(\bm k)$ for finite $k>0$. In contrast in all other cases where $\bm \gamma\neq \gamma{\hat z}$ - that is the cases where $\bm \gamma$ has non-zero component perpendicular to ${\hat z}$ or obeys $\bm \gamma=-\gamma{\hat z}$ like in convection - there are exponentially growing modes of Eqs.~(\ref{o3}).

In the case of convection the described growing modes correspond to the familiar Rayleigh-Benard instability. In that case the physically relevant wave-numbers obey $k^2=\pi^2/d^2+k_{\perp}^2$ (remind that $d$ is the distance between the plates), $k_{\perp}^2$ is arbitrary (we consider free-free boundary conditions) \cite{Drazin}. One has
\begin{eqnarray}&&
\lambda=\frac{1}{2}\biggl[\sqrt{(\nu-\kappa)^2(\pi^2/l_d^2+k_{\perp}^2)^2+\frac{4g \gamma k_{\perp}^2}{(\pi^2/l_d^2+k_{\perp}^2)}}
\nonumber\\&&-(\nu+\kappa)\left(\pi^2/l_d^2+k_{\perp}^2\right)\biggr].
\end{eqnarray}
This is negative both at small and large $k_{\perp}^2$ with maximum of $\lambda(k_{\perp})$ reached at $0<k_{\perp}<\infty$. This maximum is positive (instability) if $g\gamma/\nu\kappa>27\pi^4/4l_d^4$ and negative otherwise (stability). This is the well-known criterion where $g\gamma l_d^4/\nu\kappa$ is the Rayleigh number \cite{Drazin}. Similar instability criterion can be derived for other directions of $\bm \gamma$ (since these have no direct physical significance we do not do this).

Returning to the case of arbitrary smooth Boussinesq flow we observe that the wave-numbers with $k^4<g(\gamma-\gamma_z)/2\nu\kappa$ are permissible provided the corresponding wave-length
is much smaller than the minimum of $l_{dif}$ and $l_{\nu}$ which is the condition of validity of our study. Introducing the length $L=(\nu\kappa/g\gamma)^{1/4}$ we find the condition $L\ll l_d(1-\gamma_z/\gamma)^{1/4}$. The characteristic value of the growth exponent is $\lambda(k=0)\sim \sqrt{g(\gamma-\gamma_z)}$.

If $\mathrm{Pr}\gtrsim 1$ then  $t_{dif}=\lambda^{-1}$, $\min[l_{dif}, l_{\nu}]=l_{dif}$ and the wave-numbers with $k^4<g[\gamma-\gamma_z]/[2\nu\kappa]$ are permissible provided $l_{dif}(1-\gamma_z/\gamma)^{1/4}/L\gg 1$. If the latter condition is met then $\lambda(k=0)t_{dif}=(1-\gamma_z/\gamma)^{1/2} l_{dif}l_{\nu}/L^2\gg 1$.

If $\mathrm{Pr}\ll 1$ then $t_{dif}=\lambda^{-1}\mathrm{Pr}^{-1/2}$, $\min[l_{dif}, l_{\nu}]=l_{\nu}$ and the wave-numbers with $k^4<g(\gamma-\gamma_z)/2\nu\kappa$ are permissible provided $l_{\nu} (1-\gamma_z/\gamma)^{1/4}/L\gg 1$. If this condition is obeyed then again $\lambda(k=0)t_{dif}=(1-\gamma_z/\gamma)^{1/2}\mathrm{Pr}^{-1} l^2_{\nu}/L^2$ is much greater than one.

We conclude that if the wave-numbers with $k^4<g(\gamma-\gamma_z)/2\nu\kappa$ are permissible, then they grow exponentially with the typical time-scale much smaller than $t_{dif}$.

We are now ready to deal with the time-dependence of $\bm \gamma(t)$: because the growth occurs at the time scales much smaller than the time-scale of variations of $\bm \gamma$ then the growth occurs at the rate $\lambda(\bm k)$ determined by the instantaneous value of $\bm \gamma(t)$. This can be proved by noting that multiplying Eq.~(\ref{o1}) with time-dependent $\bm \gamma$ one still finds that Eq.~(\ref{o2}) holds approximately. This is because of inequalities like $\bm \gamma \partial_t \bm u'\gg \bm u'\partial_t\bm \gamma$. Finally, Eq.~(\ref{o2}) still has the same exponential solutions because the
time-derivatives of $\lambda(\bm k)$ can be neglected similarly.

We conclude that arbitrary smooth Bousinesq flow is unstable with respect to small-scale perturbations if $L\ll l_d(1-\gamma_z/\gamma)^{1/4}$. We stress that this conclusion is reached by finding deterministic solution though the flow itself can be turbulent. This has non-trivial consequences for flows realizable in nature where small perturbations are present always.
It implies that there are no stationary turbulent flows or time-independent laminar flows with $L\ll l_d(1-\gamma_z/\gamma)^{1/4}$. We now consider the applications of this stability condition.

\section{Instability of stable stratification in the limit of large $\mathrm{Ra}$}

In this Section we demonstrate that in the limit of large Rayleigh numbers "stably stratified fluid" will necessarily become unstable in natural situations. This is despite that infinitesimal perturbations of the stably stratified state decay.

In the case of stably stratified fluid at rest the scale $l_d$ is the scale $l_{dif}$ of variations of temperature. The ratio $l_d^4/L^4$ then defines the usual Rayleigh number $\mathrm{Ra}$.

In natural situations there are always disturbances that produce certain finite (constant or not) level of fluctuations of horizontal components of $\bm \gamma$. Then the criterion derived in the previous Section says that these fluctuations have to obey $\mathrm{Ra}^{1/4}(1-\gamma_z/\gamma)^{1/4}\lesssim 1$ in order for the fluid to remain at rest. Due to the smallness of exponent $1/4$ the factor $(1-\gamma_z/\gamma)^{1/4}$ would be close to unity for rather small fluctuations of $\bm \gamma$. It follows that in the limit of large $\mathrm{Ra}$ taken at constant level of fluctuations of $1-\gamma_z/\gamma$ the fluid would always turn unstable.

This conclusion implies that despite the linear stability of the stratified fluid at rest in the limit $\mathrm{Ra}\to\infty$ the instability would have to occur because strictly vertical $\bm \gamma$ is impossible. This probably is the reason why stably stratified fluid at rest seems to be rather rare in nature. We conjecture that this instability is one of the reasons for the peculiar features of the distribution of temperature in the ocean \cite{thorpe05} leaving the corresponding study to future work.

We now consider the implications of the instability to the turbulent flow.

\section{Implications of small-scale instability for turbulence} \label{turbulence}

In the turbulent flow the fluctuations of horizontal components of $\bm \gamma$ are non-small so that $(1-\gamma_z/\gamma)^{1/4}\sim 1$ and stability demands $L\gtrsim min[l_{dif}, l_{\nu}]$. This is compatible with the phenomenological relations proposed previously. We have \cite{Lohse,Shr}
\begin{eqnarray}&&
\epsilon=\frac{\nu^3}{L_0^4}(\mathrm{Nu}-1)\mathrm{Ra} \mathrm{Pr}^{-2},\ \ \gamma_0=\frac{\Delta}{L_0}\sqrt{\mathrm{Nu}}, \label{known}
\end{eqnarray}
where $\Delta$ is the temperature difference at the outer scale $L_0$, $\mathrm{Ra}=g\Delta L_0^3/(\nu\kappa)=L_0^4/(L^4\sqrt{\mathrm{Nu}})$. We defined the typical value of $\nabla\phi$ by $\gamma_0=\sqrt{\langle[\nabla\phi]^2\rangle}$. The last of the equations (\ref{known}) defines the Nusselt number. Since $\nabla\phi$ is due to small scale turbulence with scale $l_{dif}$ that is typically much smaller than $L_0$ then $\mathrm{Nu}\gg 1$. The relations (\ref{known}) imply
\begin{eqnarray}&&
L^4=\frac{\kappa^2}{\lambda^2}\sqrt{\mathrm{Nu}}.\nonumber
\end{eqnarray}
If $\mathrm{Pr}\gtrsim 1$ then $l_d=min[l_{dif}, l_{\nu}]=l_{dif}=\sqrt{\kappa/\lambda}$. The above relation implies then that $L^4=l_d^4\sqrt{\mathrm{Nu}}\gtrsim l_d^4$. If $\mathrm{Pr}\ll 1$ then $min[l_d, l_{\nu}]=l_{\nu}$ and $L^4=\mathrm{Pr}^{-2}l_{\nu}^4 \sqrt{\mathrm{Nu}}$. In both cases $L\gtrsim min[l_{dif}, l_{\nu}]$ is obeyed. We conclude that the stability condition $L\gtrsim min[l_d, l_{\nu}]$ is consistent with the phenomenological relations (\ref{known}).

Finally, it is worth to emphasize that the mean field description of the large-scale flow fails in describing the behavior of small-scale perturbations. Substituting into the equations the average $\bm \gamma$ which is vertical would lead to decay law for perturbations in contrast to the growth that holds for time-dependent $\bm \gamma(t)$ which is non-vertical. The solution described previously holding for given realization of the flow has to be used.

\section{Small particle as constant source of perturbations}

We now consider the problem of describing the perturbation flow around small particle translating in the Bousinessq flow. Focusing on the case of small particles of the sizes much smaller than the scales of the flow we will see that in the near vicinity of the particles the perturbation flow is described by the usual Stokes flow (see the end of the Section). Our interest thus is in the flow far from the particle. Considering scales much larger than the particle's size, the impact of the particle on the balance of momentum and energy can be described as point force with magnitude equal to minus the force that the flow exerts on the particle. The latter is the Stokes force since the force is determined by the flow near the particle's surface that is close to the usual Stokes flow. The Navier-Stokes equations become
\begin{eqnarray}&&\!\!\!\!\!\!\!\!\!\!\!\!\!\!\!\!\!
\partial_t\bm v+\bm v\cdot\nabla\bm v=-\nabla p_0+\phi \bm g+\nu\nabla^2\bm v+f{\hat z}\delta[\bm x-\bm y(t)],\label{ns}
\end{eqnarray}
where $\bm y(t)$ is the position of the particle, $|f|=6\pi \mu a |\dot {\bm y}|$ where $a$ is the particle's radius (we consider spherical particles for clarity) and it is implied that there is force driving the flow. We consider the case of particle moving in $z-$direction with the sign of $f$ determined by whether the motion is upward or downward. The cases of transversal motion can be studied similarly to the study below using the superposition. The particle's coordinate obeys
\begin{eqnarray}&&\!\!\!\!\!\!\!\!\!\!\!\!\!\!\!\!\!
\frac{d^2\bm y}{dt^2}=-\frac{\dot {\bm y}-\bm u[t, \bm y(t)]}{\tau},
\end{eqnarray}
where $\tau$ is the Stokes relaxation time given by the ratio of the particle's mass to $6\pi \mu a$ and $\bm u$ is the flow that would hold without the particle.

The particle could exchange heat with the flow. That would produce point source in the equation on $\phi$ in Eqs.~(\ref{basic0}) which could lead to non-trivial type of flow. In this work we will not consider this possibility postponing it to future work: due to linearity of equations on the perturbation flow (see below) this possible heat source term in the equation on $\phi$ can be included by superposition. Here we use the evolution of $\phi$
\begin{eqnarray}&&
\partial_t\phi+\bm v\cdot\nabla\phi=\kappa\nabla^2\phi.
\end{eqnarray}
Since the Stokes force is proportional to the particle's radius then the force term in Eq.~(\ref{ns}) is small for small particles. Thus the perturbation flow produced by the particles will be considered using linearized equations.

We decompose the flow into the background turbulent flow and the perturbation flow centered at the particle's location,
$\bm v(\bm x)=\bm u(\bm x)+\bm w'\left[\bm x-\bm y(t)\right]$, $p/\rho(\bm x)=P_0(\bm x)+P\left[\bm x-\bm y(t)\right]$ and $\theta(\bm x)=\theta_0(\bm x)+\Theta\left[\bm x-\bm y(t)\right]$, where $\bm w'$, $P$, $\Theta$ decay at large $\bm r\equiv\bm x-\bm y(t)$. The perturbations obey, cf. \cite{MaxeyRiley},
\begin{eqnarray}&&\!\!\!\!\!\!\!
\partial_t\bm w'+\left[\bm u\left(\bm x+\bm y\left[t\right]\right)-\dot {\bm y}\right]\cdot \nabla \bm w'+\left(\bm w'\cdot\nabla\right)\bm u\left(\bm x+\bm y\left[t\right]\right)
\nonumber\\&&\!\!\!\!\!\!\!+\bm w'\cdot\nabla\bm w'
=-\nabla P+\Theta \bm g +\nu\nabla^2\bm w'+f{\hat z}\delta(\bm x),\ \ \ \
\nonumber\\&&\!\!\!\!\!\!\!\partial_t\Theta+\bm w'\cdot\nabla\theta_0\left(\bm x+\bm y\left[t\right]\right)
+\left[\bm u\left(\bm x+\bm y\left[t\right]\right)-\dot {\bm y}\right]\cdot\nabla\Theta\nonumber\\&&\!\!\!\!\!\!\!
+\bm w'\cdot\nabla\Theta=\kappa\nabla^2\Theta
,\ \ \nabla\cdot\bm w'=0.\label{a01}
\end{eqnarray}
For small particles the Reynolds number $\mathrm{Re_p}=w'_da/\nu$ and the P\'eclet number $\mathrm{Pe}_p=w'_da/\kappa$ based on the particle's drift velocity with respect to the flow  $\bm w'_d=\dot {\bm y}-\bm u[t, \bm y(t)]$ are small. In particular when the Stokes number $\mathrm{St}=\lambda \tau\ll 1$ the drift velocity is given by \cite{Maxey}
\begin{eqnarray}&&\!\!\!\!\!\!\!
\bm w'_d=-\tau\left[\partial_t \bm u+(\bm u\cdot\nabla)\bm u\right].
\end{eqnarray}
It follows that $w'\sim \mathrm{St} \lambda l_{\nu}$ so that $\mathrm{Re_p}=\mathrm{St} a/l_{\nu}$, $\mathrm{Pe_p}=\mathrm{Re_p} \mathrm{Pr}$. Thus for small particles one can neglect the non-linear terms in Eqs.~(\ref{a01}). This quantifies the smallness of $f$ needed for the perturbation flow to obey linearized equations. Further we observe that due to smallness of $\mathrm{Re_p}$, $\mathrm{Pe_p}$ one can neglect the $\bm w'_d$ term in $\bm u\left(\bm x+\bm y\left[t\right]\right)-\dot {\bm y}=\bm u\left(\bm x+\bm y\left[t\right]\right)-\bm u[t, \bm y(t)]-\bm w'_d$ in comparison with viscous and heat conductance terms. Focusing on scales much smaller than $l_{\nu}$ one finds $\bm u\left(\bm x+\bm y\left[t\right]\right)-\bm u[t, \bm y(t)]=\sigma\bm x$ where $\sigma_{ik}(t)$ is the matrix of velocity gradients at the position of the particle $\sigma_{ik}(t)=\nabla_k u_i[t, \bm y(t)]$. Thus, Eqs.~(\ref{a01}) become
\begin{eqnarray}&&\!\!\!
\partial_t\bm w'=-\nabla P+\Theta \bm g+\nu\nabla^2\bm w'+f{\hat z}\delta(\bm x),\nonumber\\&&\!\!\!
\partial_t\Theta+\bm w'\cdot\nabla\theta_0\left(\bm x+\bm y\left[t\right]\right)+\sigma\bm x\cdot \nabla\Theta=\kappa\nabla^2\Theta. \label{start}
\end{eqnarray}
where we neglected $\sigma\bm x\cdot \nabla \bm w'$, $\sigma \bm w'$ terms in the equation on $\bm w'$ in comparison with the viscous terms using that we consider the flow at scales much smaller than $l_{\nu}$. We did not however neglect the $\sigma\bm x\cdot \nabla\Theta$ term in the equation on $\Theta$ since it is possible that this term is not small if Prandtl number is large: e.g., in the ocean the typical values are $\mathrm{Pr}\approx 7$ for temperature and $\approx 670$ for salinity \cite{thorpe05}.

We observe from Eq.~(\ref{start}) that $\theta_r$ at scale $r$ obeys $\theta_r\sim r^2\gamma w'_r/\kappa$ where $\gamma$ is the typical value of $\nabla\theta_0$ and we balance the heat conductance term with the term $\bm w'\cdot \nabla\theta_0$. It follows that in the first of Eqs.~(\ref{start}) we have at scale $r$
\begin{eqnarray}&&\!\!\!
\frac{\Theta_r g}{\nu\nabla^2 w'_r}\sim \frac{r^4}{L^4}.
\end{eqnarray}
Thus buoyancy term is negligible at scales $r^4\ll L^4$ where one would have the usual Stokeslet flow unless $l_d$ is smaller than $L$. The Stokes flow that holds near the particle will get modified at the smallest of the scales $L$, $l_d$ (the Oseen radius is assumed to be larger than these scales).
We start with the case $L\ll l_d$. The condition of stability in this case brings us to consider the following range of parameters.

\section{Flow around small particle at $l_d\gg L\gtrsim l_d(1-\gamma_z/\gamma)^{1/4}$: quasi-exponential decay}

We consider flow around the particle at the scale $L$ when we deal with weakly fluctuating stratification $L\ll l_d$, but $L\gtrsim l_d(1-\gamma_z/\gamma)^{1/4}$. The inequalities imply that $(1-\gamma_z/\gamma)^{1/4}\ll 1$ so that only small fluctuations of $\bm \gamma$ around the stratification relation $\gamma=\gamma_z$ hold in the flow.
Due to $L\ll l_d$ one can neglect $\sigma\bm x$ term in the equation (\ref{start}) and put $\nabla\theta_0\left(\bm x+\bm y\left[t\right]\right)\approx -\bm \gamma(t)$ where $\bm \gamma(t)=-\nabla\theta_0\left(\bm y\left[t\right]\right)$. Here we assume that the scale $l_{dif}$ of spatial variations of $\nabla\theta_0$ is much larger than the scale of consideration $L$. We find
\begin{eqnarray}&&\!\!\!
\partial_t\bm w'=-\nabla P+\Theta \bm g+\nu\nabla^2\bm w'+f{\hat z}\delta(\bm x),\\&&\!\!\!
\partial_t\Theta=\bm \gamma\cdot\bm w'+\kappa\nabla^2\Theta.
\end{eqnarray}
We thus obtained Eqs.~(\ref{sist}) governing small-scale fluctuations that are created by the constant momentum source $f$. Clearly, in the case of stable stratification $\bm \gamma=\gamma{\hat z}$ the condition of stability of the flow $L\gtrsim l_d(1-\gamma_z/\gamma)^{1/4}$ implies that linear modes of the above equation decay. Thus in the presence of the constant source there is time-independent solution determined from
\begin{eqnarray}&&\!\!\!
0=-\nabla P+\Theta \bm g+\nu\nabla^2\bm w'+f{\hat z}\delta(\bm x),\label{s1}\\&&\!\!\!
0=\bm \gamma\cdot\bm w'+\kappa\nabla^2\Theta. \label{s2}
\end{eqnarray}
This equation in the case of stable stratification $\bm \gamma=\gamma{\hat z}$ was considered in \cite{Ardekani}. It is useful for what comes later to keep $\bm \gamma$ arbitrary. The equation is solved by the Fourier transform
\begin{eqnarray}&&
\!\!\!\!\!\!\!\!i\bm k P=\Theta \bm g -\nu k^2\bm w'+f{\hat z},\ \ \bm k\cdot\bm w'=0,\\&&
\!\!\!\!\!\!\!\!\bm \gamma\cdot \bm w'=\kappa k^2\Theta.
\end{eqnarray}
We multiply the first equation with $\bm k$ and use the incompressibility condition $\bm k\cdot\bm w'=0$ to eliminate the pressure,
\begin{eqnarray}&&
P=i g k_z\Theta'/k^2,\ \ \Theta'\equiv\Theta-f/g. \label{primed}
\end{eqnarray}
Introducing ${\hat k}\equiv \bm k/k$ and the projection $\Pi_{ij}(\bm k)$ leads to
\begin{eqnarray}&&
\!\!\!\!\!\!\!\!\nu k^2\bm w'= \Theta' \Pi(\bm k)\bm g,\ \
\Pi_{ij}(\bm k)=\delta_{ij}-{\hat k}_i{\hat k}_j
.\label{a15}
\end{eqnarray}
It follows that
\begin{eqnarray}&&
\bm \gamma \cdot \bm w'=-\frac{g\Theta'{\tilde \gamma}}{\nu k^2},\label{vel}
\end{eqnarray}
where ${\tilde \gamma}(\bm k)=\gamma_z-(\bm \gamma \cdot{\hat k}) {\hat k}_z$ was introduced previously.
Substituting $\bm \gamma \cdot \bm w'$ in the equation for $\Theta$,
\begin{eqnarray}&&\!\!\!\!\!\!\!\!
\Theta=\frac{\phi(\bm k)}{\alpha(\bm k)}=\frac{f}{g}\frac{{\tilde \gamma}/|\gamma_z|}{L^4 k^4+{\tilde \gamma}/|\gamma_z|}=\frac{f}{g}-\frac{f}{g}\frac{L^4 k^4}{L^4 k^4+{\tilde \gamma}/|\gamma_z|}.\nonumber \nonumber\\&& \!\!\!\!\!\!\!\!
 \alpha(\bm k)\equiv \kappa k^2d(\bm k),\ \ d(\bm k)\equiv
1+\frac{{\tilde \gamma}/|\gamma_z|}{L^4 k^4},\ \ 
\phi(\bm k)\equiv \frac{f{\tilde \gamma}}{\nu k^2},\nonumber
\end{eqnarray}
Thus
\begin{eqnarray}
&&\Theta'=-\frac{f}{g}\frac{L^4 k^4}{L^4 k^4+{\tilde \gamma}/|\gamma_z|}. \label{b12}\!\!\!\!\!\!\!\!
\end{eqnarray}
This solution in the case of stable stratification $\bm \gamma=\gamma{\hat z}$ where one has
\begin{eqnarray}
&&\Theta'=-\frac{f}{g}\frac{L^4 k^4}{L^4 k^4+k_{\perp}^2/k^2}, \!\!\!\!\!\!\!\!\label{solution}
\end{eqnarray}
was obtained in \cite{Ardekani}. It was demonstrated using this formula that the flow at scales $L\lesssim r\ll l_b$ is profoundly altered by the buoyancy force. Here $l_b$ is depth of the fluid layer or the characteristic scale of variations of $\bm \gamma$. Instead of the typical Stokesian $1/r$ behavior, fast decaying velocity was found. This observation was done numerically with no explanation of the nature of the decay (power-law with large exponent, exponential decay or other). We pass to provide quantitative consideration of the nature of the decay.

The flow is axially symmetric, so that one can introduce the stream function  $\psi_0(r, z)=\int_0^r r' w_z'(r', z)dr'$ that obeys in cylindrical coordinates
\begin{eqnarray}&&
w_z'=\frac{1}{r}\frac{\partial \psi_0}{\partial r},\ \ w_r'=-\frac{1}{r}\frac{\partial \psi_0}{\partial z}.
\end{eqnarray}
We have (it is useful to introduce averaging over $\varphi$)
\begin{eqnarray}&&\!\!\!\!\!\!\!\!\!
\psi_0\!=\!\!\int_0^r\! \int_0^{2\pi}\!\frac{r'dr'd\varphi}{2\pi} \!\int\!  \frac{w_z'(\bm k)d\bm k}{(2\pi)^3}\exp\left[ik_z z\!+\!ik_{\perp}r'\cos\varphi\right]\nonumber\\&&\!\!\!\!\!\!\!\!\!
=-\frac{g}{\nu}\int \frac{k_{\perp}^2 d\bm k}{(2\pi)^3 k^4}\Theta'(\bm k)\exp\left[ik_z z\right]\int_0^r r'J_0(k_{\perp}r')dr',\nonumber
\end{eqnarray}
where we used $w'_z=-gk_{\perp}^2\Theta'(\bm k)/\nu k^4$ and $J_0(x)$ is the Bessel function of zeroth order.
Noting that $k_{\perp}^2 \int_0^r r'J_0(k_{\perp}r')dr'= \int_0^{k_{\perp} r} xJ_0(x)dx=k_{\perp}rJ_1(k_{\perp}r)$, passing to the dimensionless
integration variable $\bm q=L \bm k$ and scaling all lengths (i.e. $r\:,z$)  with $L$,
\begin{eqnarray}&&\!\!\!\!\!\!\!
\psi_0(L\bm r)=\frac{f L r}{\nu}\int_0^{\infty} \frac{q_{\perp}^2dq_{\perp}}{(2\pi )^2}\int_{-\infty}^{\infty} \frac{\exp\left[iq_z z\right]J_1(q_{\perp}r)q^2 dq_z}{q_{\perp}^2+q^6}
\nonumber,
\end{eqnarray}
where we used the solution (\ref{solution}). To further simplify this expression we introduce polar coordinates in $(q_z, q_{\perp})$ plane by $q_{\perp}=q\sin\theta$, $q_z=q \cos\theta$, so that
\begin{eqnarray}&&\!\!\!\!\!\!\!\!
\psi_0(L\bm r)\!=\!\frac{f L r}{\nu}\!\int_0^{\pi/2}\! \frac{\sin^2\theta d\theta}{2\pi^2}\int_0^{\infty}
\frac{q^3\cos\left[qz\cos\theta \right]}{\sin^2\theta+q^4} dq \nonumber\\&&\times J_1(qr\sin\theta)
\end{eqnarray}
When $q$ is large the integrand is proportional to $q^{-3/2}$ times an oscillating function of $q$, so the convergence is slow. We rewrite the integral so the convergence is fast and convenient for the numerical evaluation.
The denominator has simple poles in the upper half-plane at $q_1=\sqrt{|\sin\theta|}(1+i)/\sqrt{2}$ and  $q_2=-q_1^*$. We write
\begin{eqnarray}&&\!\!\!\!\!\!\!\!\!\!\!\!\!\!\!
\frac{1}{\sin^2\theta+q^4}=\frac{1}{(q-q_1)(q+q_1)(q-q_1^*)(q+q_1^*)}.
\end{eqnarray}
Closing the contour in the upper half-plane is not straightforward because the integrand has growing exponents when continued onto the complex plane. We first write
\begin{eqnarray}&&\!\!\!\!\!\!\!\!
\psi_0(L\bm r)\!=\!\frac{f L r}{\nu}\!\int_0^{\pi/2}\! \frac{\sin^2\theta d\theta}{2(2\pi)^2}\int_{-\infty}^{\infty}
\frac{q^3J_1(qr\sin\theta)}{\sin^2\theta+q^4} dq \nonumber\\&&\times \left(\exp\left[iqz\cos\theta \right]+\exp\left[-iqz\cos\theta \right]\right),
\end{eqnarray}
where we used that $J_1(qr\sin\theta)$ is odd function of $q$ to continue the integral over $q$ to $(-\infty, \infty)$.
Using the integral representation of $J_1(z)$ ($Re$ stands for real part)
\begin{equation}
\psi_0(L\bm r)=\frac{f Lr}{\nu}\mathrm{Re}\left\{I_0(y_+)+I_0(y_-)\right\}
\end{equation}
where $y_{\pm}=r\sin\theta\sin\phi\pm z\cos\theta$ and
\begin{eqnarray}
I_0(y)=\int_0^{\pi/2} \frac{\sin^2\theta d\theta}{2(2\pi)^2}\int_0^{\pi}\frac{d\phi}{\pi}\exp[-i\phi] && \nonumber\\
\int_{-\infty}^{\infty} \frac{q^3 dq }{\sin^2\theta+q^4} \exp\left[i q y\right] &&\nonumber
\end{eqnarray}
The integral is purely imaginary due to the parity properties of the integrand, so $I_0(y)$ is odd function of $y$. We consider $y>0$ when
we can close the contour in the upper half-plane,
\begin{eqnarray}&&
I_1\equiv\frac{1}{2\pi i} \int_{-\infty}^{\infty}\frac{q^3 dq }{\sin^2\theta+q^4} \exp\left[iqy\right]
\nonumber\\&&
=\frac{q_1^3 \exp\left[iq_1y\right]}{2q_1(q_1-q_1^*)(q_1+q_1^*)}+c. c.\nonumber\\&&
=\exp\left[-y\sqrt{\sin\theta/2}\right]\frac{\cos\left[y\sqrt{\sin\theta/2}\right]}{2}.\nonumber
\end{eqnarray}
In the case $y<0$ we close the contour in the lower half-plane which gives
\begin{eqnarray}&&
I_1=-\frac{q_1^3 \exp\left[iq_1|y|\right]}{2q_1(q_1-q_1^*)(q_1+q_1^*)}-c. c.\nonumber
\end{eqnarray}
Therefore we can write
\begin{eqnarray}&&
I_1=\mathrm{sign}(y) \exp\left[-|y|\sqrt{\sin\theta/2}\right]\frac{\cos\left[|y|\sqrt{\sin\theta/2}\right]}{2}.\nonumber
\end{eqnarray}
Thus we finally obtain
\begin{eqnarray}&&\!\!\!\!\!\!\!\!
\psi_0\!=\!\frac{f Lr}{\nu}\int_0^{\pi/2}\!\!\frac{\sin^2\theta d\theta}{(2\pi)^2}\!\int_0^{\pi/2}\!\!\!d\phi \exp\left[-h_+\right]\sin\phi\cos(h_+)s_+\nonumber\\&&\!\!\!\!\!\!\!\!
+\frac{f Lr}{\nu}\int_0^{\pi/2} \frac{\sin^2\theta d\theta}{(2\pi)^2}\int_0^{\pi/2}d\phi \exp\left[-h_-\right]\sin\phi\cos(h_-)
s_-.\nonumber
\end{eqnarray}
where $h_{\pm}\equiv |r\sin\theta\sin\phi\pm z\cos\theta|\sqrt{\sin\theta/2}$ and $s_{\pm}\equiv \mathrm{sign}(r\sin\theta\sin\phi\pm z\cos\theta)$. This form of $\psi_0$ suits well the numerical evaluation.

\begin{figure}[tbc]
\begin{center}
\includegraphics[scale=0.75]{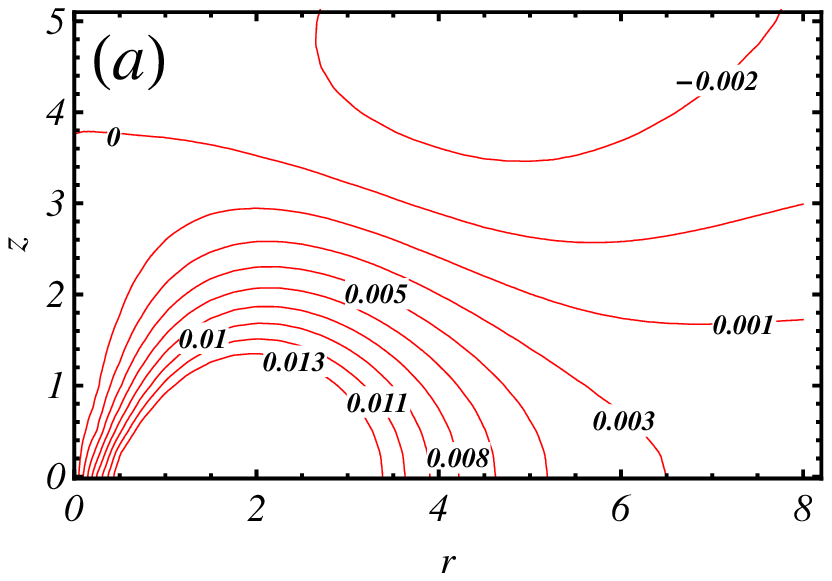} \\ 
\vskip0.1cm
\includegraphics[scale=0.8]{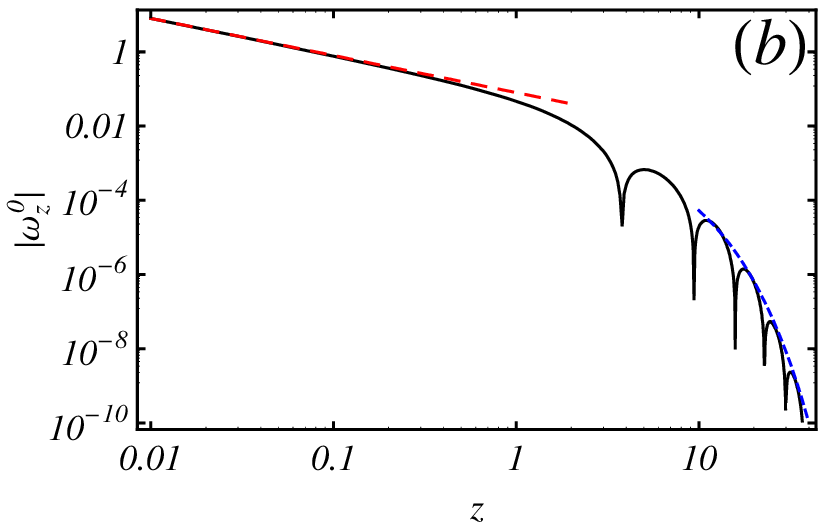}   
\end{center}
\caption{Flow induced by a vertical (upward) Stokeslet located at the origin in a vertically stratified fluid at rest (\emph{a}) streamline, the contour labels show the corresponding values of  the stream function $\psi$ (scaled with $f L/\nu$); (b) axial velocity (scaled with $f/\nu L$) vs. the axial distance $z$ (scaled with $L$):  the solid black line corresponds to $|w_z^0|$, the red dashed (long dashes) line stands for the velocity due to Stokeslet in unstratified fluid, $1/4\pi |z|$ that prevails at $z \lesssim 1$, the blue dashed line (short dashes) stands for the approximate exponential decay $|w_z^0|\propto \exp{(-y_0|z|/\sqrt{2})}$.
\label{fig:psi}}
\end{figure}

The resulting plot of the flow $\psi_0$ (scaled with $f L/\nu$) induced by a passively translating particle (Stokeslet) is shown in the Fig.~\ref{fig:psi}\emph{a}. The streamline pattern in Fig.~\ref{fig:psi}\emph{a} is the same as reported before (see Fig.~1 in \cite{Ardekani}, the axial velocity at $r=0$ vanishes at $z\simeq 3.8$).

It is instructive to consider directly the axial velocity averaged over the angles
\begin{eqnarray}&&
w_{z}^0(L\bm r)=\int_0^{2\pi}w_z'(Lr, Lz, \phi )\frac{d\phi}{2\pi}
\nonumber\\&&
=-\frac{g}{\nu L}\int \frac{q_{\perp}^2 d\bm q}{(2\pi)^3 q^4}\Theta'\left(\frac{\bm q}{L}\right)\exp\left[iq_z z\right]J_0(q_{\perp}r)\nonumber\\&&
=\frac{f}{\nu L}\int \frac{q_{\perp}^2 q^2d\bm q}{(2\pi)^3\left[q_{\perp}^2+q^6\right]}\exp\left[iq_z z\right]J_0(q_{\perp}r),\nonumber
\end{eqnarray}
where we used $w'_z=-gk_{\perp}^2\Theta'(\bm k)/\nu k^4$. Using that the integrand is even function of $q_z$,
\begin{eqnarray}&&\!\!\!\!\!\!\!\!\!
w_{z}^0=\frac{f}{\nu L}\int_0^{\infty} dq_z \int_0^{\infty} \frac{dq_{\perp}}{2\pi^2} \frac{q_{\perp}^3 q^2}{\left[q_{\perp}^2+q^6\right]}\cos\left[q_z z\right]J_0(q_{\perp}r),\nonumber
\end{eqnarray}
We consider the asymptotic forms that $w_{z}^0$ takes at large and small distances, cf. \cite{Ardekani}.
Passing to polar coordinates in $(q_{\perp}, q_z)$ plane we find
\begin{eqnarray}&&\!\!\!\!\!\!\!\!\!
{\tilde w}_0=\int_0^{\pi/2} \frac{\sin^3\theta d\theta}{(2\pi)^2} \int_{-\infty}^{\infty} dq
 \frac{q^4  \cos\left[qz\cos\theta\right] J_0(qr\sin\theta)}{\left[\sin^2\theta+q^4\right]}\nonumber\\&&\!\!\!\!\!\!\!\!\!
= \int_0^{\pi/2} \frac{\sin^3\theta d\theta}{(2\pi)^2}
\int_{-\infty}^{\infty}dq
\cos\left[qz\cos\theta\right]J_0(qr\sin\theta)\nonumber\\&&\!\!\!\!\!\!\!\!\!
-
\int_0^{\pi/2} \frac{\sin^5\theta d\theta}{(2\pi)^2}
\int_{-\infty}^{\infty} \frac{\cos\left[qz\cos\theta\right]J_0(qr\sin\theta)dq}{\left[\sin^2\theta+q^4\right]}.
\end{eqnarray}
where ${\tilde w}_0\equiv  w_z^0\nu L/f$. We note that
\begin{eqnarray}&&
\int_{-\infty}^{\infty}dq
\cos\left[qz\cos\theta\right]J_0(qr\sin\theta)=\int_0^{2\pi}d\alpha
\\&&\nonumber
\int_{-\infty}^{\infty} \frac{dq}{2\pi} \exp\left[iqz\cos\theta+iqr\sin\theta\sin\alpha\right]\\&&\nonumber
=2\int_0^{\pi/2}d\alpha \delta\left(r\sin\theta\sin\alpha-|z|\cos\theta\right).
\end{eqnarray}
One finds
\begin{eqnarray}&&\!\!\!\!\!\!\!\!\!
\int_0^{\pi/2}\! \frac{\sin^3\theta d\theta}{(2\pi)^2}
\int_{-\infty}^{\infty}\!dq\!
\cos\left[qz\cos\theta\right]J_0(qr\sin\theta)\!=\!\int_0^{\pi/2}d\alpha\! \nonumber\\&&\!\!\!\!\!\!\!\!\!
\times\int_{0}^{\pi/2} \frac{\sin^2\theta d\theta}{2\pi^2 |z|}\delta\left(\cot\theta-r\sin\alpha/|z|\right)
=\int_0^{\pi/2}\frac{d\alpha}{2\pi^2 |z|}\nonumber\\&&\!\!\!\!\!\!\!\!\!
\times\frac{1}{\left(1+r^2\sin^2\alpha/z^2\right)^2}=\frac{2+r^2/z^2}{8\pi|z|(1+r^2/z^2)^{3/2}},
\end{eqnarray}
which is nothing but the Stokeslet flow that holds without stratification. Thus we obtained the representation of the velocity as the sum of the flow without stratification and the correction due to the stratification,
\begin{eqnarray}&&
{\tilde w}_0=\frac{2+r^2/z^2}{8\pi|z|(1+r^2/z^2)^{3/2}}
-\int_0^{\pi/2} \frac{\sin^5\theta d\theta}{(2\pi)^2}
\nonumber\\&&
\times
\int_{-\infty}^{\infty} \frac{\cos\left[qz\cos\theta\right]J_0(qr\sin\theta)dq}{\left[\sin^2\theta+q^4\right]}. \label{wz0}
\end{eqnarray}
This representation is useful for studying the role of stratification. At $z\ll 1$, $r\ll 1$, one has
\begin{eqnarray}&&\!\!\!\!\!\!\!\!\!
{\tilde w}_0(rL, zL)\approx \frac{2+r^2/z^2}{8\pi|z|(1+r^2/z^2)^{3/2}}
-\int_0^{\pi/2} \frac{\sin^5\theta d\theta}{(2\pi)^2}\nonumber\\&& \!\!\!\!\!\!\!\!\!
\int_{-\infty}^{\infty} \frac{dq}{\left[\sin^2\theta+q^4\right]}\!=\!\frac{2+r^2/z^2}{8\pi|z|(1+r^2/z^2)^{3/2}}
\!-\! \frac{B(9/4, 9/4)}{\pi}\:,
\nonumber
\end{eqnarray}
where $\mathcal{B}(x, y)$ is the beta function, so that $\mathcal{B}(9/4, 9/4)/\pi \simeq0.0351$.
Thus stratification has a small impact on the flow at scales smaller than $L$, introducing a small uniform correction to the flow. It should be noted that though the correction is small, it can have finite
effect on the motion of nearby particles due to the persistent drift that it induces. The study of this drift is left for future work.

On the other hand, the stratification's contribution is dominant at scales larger than $L$, screening the Stokeslet flow, so that the resulting flow is fully determined by the stratification. To demonstrate this we consider the axial velocity along the axis of symmetry $r=0$,
\begin{eqnarray}&&\!\!\!\!\!\!\!\!\!
{\tilde w}_0(0, zL)
\!=\!\frac{1}{4\pi |z|}
\!-\!
\int_0^{\pi/2} \! \frac{\sin^5\theta d\theta}{(2\pi)^2}\!
\int_{-\infty}^{\infty}\! \!\frac{\exp\left[iq|z|\cos\theta\right]\!dq}{\left[\sin^2\theta+q^4\right]}.\nonumber
\end{eqnarray}
We note that
\begin{eqnarray}&&
\frac{1}{2\pi i}\int_{-\infty}^{\infty} \frac{\exp\left[iq|z|\cos\theta\right]dq}{\left[\sin^2\theta+q^4\right]}=\frac{\exp\left[iq_1|z|\cos\theta\right]}{2q_1(q_1^2-q_1^{*2})}-c. c.\nonumber\\&&
=\frac{\exp\left[iq_1|z|\cos\theta-i\pi/4\right]}{4i \sin^{3/2}\theta}-c. c.
=\frac{1}{2i\sin^{3/2}\theta}
\\&&\times
\exp\left[-|z|\cos\theta\sqrt{\frac{\sin\theta}{2}}\right]
\cos\left[|z|\cos\theta\sqrt{\frac{\sin\theta}{2}}-\frac{\pi}{4}\right].\nonumber
\end{eqnarray}
Thus, introducing $\varphi=\pi/2-\theta$,
\begin{eqnarray}&&\!\!\!\!\!\!\!\!\!
{\tilde w}_0(r=0)
=\frac{1}{4\pi |z|}
-
\int_0^{\pi/2} \cos\left[|z|\sin\varphi\sqrt{\frac{\cos\varphi}{2}}-\frac{\pi}{4}\right]\nonumber\\&&
\exp\left[-|z|\sin\varphi\sqrt{\frac{\cos\varphi}{2}}\right]
\frac{\cos^{7/2}\varphi d\varphi}{4\pi}
.\label{skor}
\end{eqnarray}
When $|z|$ is large the integral is determined by the minima of $\sin\varphi\sqrt{\cos\varphi/2}$ which are equal to zero and attained at $\varphi=0$ and $\varphi=\pi/2$. The contribution of the saddle-point $\varphi_0$ where $\sin\varphi\sqrt{\cos\varphi}$ has zero derivative is to be considered too, though it includes the exponentially small factor, see below.
The contribution of the
leading order term that comes from the neighborhood of $\varphi=0$ is
\begin{eqnarray}&&\!\!\!\!\!\!\!\!\!
\int_0^{\infty} \cos\left[|z|\varphi/\sqrt{2}-\pi/4\right]
\exp\left[-|z|\varphi/\sqrt{2}\right]
\frac{d\varphi}{4\pi}\nonumber\\&&
=Re \frac{\sqrt{2}\exp[i\pi/4]}{4\pi |z|(1+i)}=\frac{1}{4\pi |z|},
\end{eqnarray}
demonstrating that stratification screens the Stokeslet solution at $|z|\gg L$. The contribution of the
leading order term coming from the neighborhood of $\varphi=\pi/2$ is
\begin{eqnarray}&&\!\!\!\!\!\!\!\!\!
\int_0^{\pi/2} \cos\left[|z|\sqrt{(\pi/2-\varphi)/2}-\pi/4\right]\label{nbhd}\\&&
\exp\left[-|z|\sqrt{(\pi/2-\varphi)/2}\right]
\frac{(\pi/2-\varphi)^{7/2}d\varphi}{4\pi}\propto \frac{1}{|z|^9}\:.\nonumber
\end{eqnarray}
We conclude that the velocity field decays faster than the Stokeslet solution at large distances from the particle.
To find the leading order term at large $|z|$ we split the domain of integration over $\varphi$ into $\varphi<\varphi_0$ and $\varphi>\varphi_0$, where
$\varphi_0\equiv \arctan\sqrt{2}$,
\begin{eqnarray}&&\!\!\!\!\!\!
{\tilde w}_0(r=0)
=\frac{1}{4\pi |z|}
-
\int_0^{\varphi_0} \cos\left[|z|\sin\varphi\sqrt{\cos\varphi/2}-\pi/4\right]\nonumber\\&& \!\!\!\!\!\!
\frac{\exp\left[-|z|\sin\varphi\sqrt{\cos\varphi/2}\right]\cos^{7/2}\varphi d\varphi}{4\pi}\!-\!\int_0^{\pi/2-\varphi_0}\!\frac{\sin^{7/2}{\tilde \varphi} }{4\pi}
\nonumber\\&& \!\!\!\!\!\!
\cos\left[|z|\cos{\tilde \varphi}\sqrt{\sin{\tilde \varphi}/2}\!-\!\pi/4\right]
\exp\left[-|z|\cos{\tilde \varphi}\sqrt{\sin{\tilde \varphi}/2}\right]d{\tilde \varphi},\nonumber
\end{eqnarray}
where 
we introduced ${\tilde \varphi}=\pi/2-\varphi$. Then,
in the first of the integrals, designated by $I'$, one can pass to the integration variable
$y\equiv \sin\varphi\sqrt{\cos\varphi}$,
\begin{eqnarray}&&
\frac{d\varphi}{dy}=\frac{2\sqrt{\cos\varphi}}{2\cos^{2}\varphi-\sin^2\varphi}.
\end{eqnarray}
Introducing $y_0\equiv y\left(\varphi_0\right)=\sqrt{2/(3\sqrt{3})}$ (we use $\cos[\arctan(\sqrt{2})]=1/\sqrt{3}$) one finds
\begin{eqnarray}&&\!\!\!\!\!\!
I'\!=\!Re\int_0^{y_0}\!h(y) \exp\left[-\frac{(1+i)|z|y}{\sqrt{2}}+\frac{i\pi}{4}\right] \frac{dy}{2\pi},\nonumber
\end{eqnarray}
where we defined
\begin{eqnarray}&&\!\!\!\!\!\!\!\!\!\!\!\!
h(y)=\frac{x_1^4(y)}{3x_1^2(y)-1},\ \  y=\sqrt{x_1(y)-x_1^3(y)}\label{o1}
\end{eqnarray}
with $x_1=\cos\varphi(y)$, so that $x_1(y)$ is the branch of the solution of the cubic equation $x^3-x=-y^2$ that obeys $x(0)=1$. One has
\begin{eqnarray}&&\!\!\!\!\!\!\!\!\!\!\!\!\!\!\!
\frac{dx}{dy}=\frac{2y}{1-3x^2},\ \ \frac{dh}{dy}=\frac{4x^3\left(3x^2-2\right)y}{\left(1-3x^2\right)^3}.
\end{eqnarray}
The large $|z|$ expansion is obtained by introducing the Taylor expansion of $h(y)$ into $I'$,
\begin{eqnarray}&&\!\!\!\!\!\!
I'\!\approx \!\sum_{n=0}^{\infty} \frac{h^{(2n)}(0)}{(2n)!}Re\int_0^{\infty}\! y^{2n} \exp\left[-\frac{(1+i)|z|y}{\sqrt{2}}+\frac{i\pi}{4}\right] \frac{dy}{2\pi},\nonumber\\&&\!\!\!\!\!\!
=\sum_{n=0}^{\infty}h^{(2n)}(0)Re\frac{1}{2 i^n|z|^{2n+1}\pi}=\sum_{k=0}^{\infty}\frac{(-1)^k h^{(4k)}(0)}{2|z|^{4k+1}\pi},
\end{eqnarray}
where we used that $h^{(2n)}(0)=0$ by $h(y)=h(-y)$. The $k=0$ term reproduces $(4\pi |z|)^{-1}$ found previously. The $k=1$ term turns out to be vanishing because a direct computation reveals that $h^{(4)}(0)=0$. The next order term is proportional to $|z|^{-9}$, which is the same order as in Eq.~(\ref{nbhd}), necessitating the consideration of the contribution of the neighborhood of $\varphi=\pi/2$ into ${\tilde w}_0$. This is described by
\begin{eqnarray}&&\!\!\!\!\!\!
I''\equiv Re \int_0^{\pi/2-\varphi_0}\!\frac{\sin^{7/2}{\tilde \varphi} }{4\pi}\exp\left[-\frac{(1+i)|z|{\tilde y}}{\sqrt{2}}+\frac{i\pi}{4}\right]d{\tilde \varphi},\nonumber
\end{eqnarray}
where ${\tilde y}=\cos{\tilde \varphi}\sqrt{\sin{\tilde \varphi}}$. Passing to the integration variable ${\tilde y}$, using
\begin{eqnarray}&&
\frac{d\tilde \varphi}{dy}=\frac{2\sqrt{\sin\tilde \varphi}}{\cos^2\tilde \varphi-2\sin^{2}\tilde \varphi}.
\end{eqnarray}
we obtain
\begin{eqnarray}&&\!\!\!\!\!\!
I''\!=-Re\int_0^{y_0}\!{\tilde h}({\tilde y}) \exp\left[-\frac{(1+i)|z|{\tilde y}}{\sqrt{2}}+\frac{i\pi}{4}\right] \frac{d{\tilde y}}{2\pi},\nonumber
\end{eqnarray}
where we defined
\begin{eqnarray}&&\!\!\!\!\!\!\!\!\!\!\!\!
{\tilde h}({\tilde y})=\frac{x_2^4({\tilde y})}{3x_2^2({\tilde y})-1},\ \  y=\sqrt{x_2(y)-x_2^3(y)}\nonumber
\end{eqnarray}
with $x_2({\tilde y})=\sin{\tilde \varphi}({\tilde y})$ given by the branch of the solution of the cubic equation ${\tilde x}^3-{\tilde x}=-{\tilde y}^2$ that obeys $x(0)=0$.
Combining $I'$ and $I''$ the following representation of ${\tilde w}_0$ is obtained
\begin{eqnarray}&&\!\!\!\!\!\!
{\tilde w}_0
=\frac{1}{4\pi |z|}
-\int_0^{y_0}\!\biggl[\frac{x_1^4(y)}{3x_1^2(y)-1}-\frac{x_2^4(y)}{3x_2^2(y)-1}\biggr]\\&&
\exp\left[-\frac{|z|y}{\sqrt{2}}\right]\cos\left(\frac{|z|y}{\sqrt{2}}-\frac{\pi}{4}\right) \frac{dy}{2\pi}.\label{wz00}
\end{eqnarray}
Introducing
\begin{eqnarray}&&\!\!\!\!\!\!
l(y)\equiv \frac{x_1^4(y)}{3x_1^2(y)-1}-\frac{x_2^4(y)}{3x_2^2(y)-1},
\end{eqnarray}
we find the asymptotic series for ${\tilde w}_0$ at large $|z|$,
\begin{eqnarray}&&\!\!\!\!\!\!
{\tilde w}_0\!= -\sum_{k=2}^{\infty}\frac{(-1)^k l^{(4k)}(0)}{2|z|^{4k+1}\pi}.
\end{eqnarray}
Thus at very large $|z|$ the axial velocity $\tilde{w}_0(r=0)$ is expected to decay as $\sim |z|^{-9}$, as $l^{(8)}(0)\neq 0$. The absolute value of the axial velocity at $r=0$ determined by numerical integration of Eq.~\ref{wz00} is shown in Fig.~\ref{fig:psi}\emph{b}.
It agrees with the earlier results in \cite{Ardekani} and shows that at length scales below $L$ the flow is just that due to unstratified Stokeslet solution $\sim 1/|z|$ (dashed red line in Fig.~\ref{fig:psi}\emph{b}). At scales $\gtrsim L$ the Stokeslet flow is screened by the buoyant flux due to vertical stratification resulting in a series of eddies with velocity decaying much faster than $\sim 1/|z|$. The numerical results suggest that at large, but finite $|z|$, the saddle-point contribution dominates the integral in (\ref{wz00}) so that the velocity decays exponentially fast $\propto \exp{(-y_0|z|/\sqrt{2})}$ (dashed blue line in Fig.~\ref{fig:psi}\emph{b}), and not $\propto |z|^{-9}$ as was suggested above. It seems that the involved numbers are such that the power-law will be seen only at very large $|z|$ when the solution is vanishingly small. For practical purposes, therefore, one can say that the velocity decays exponentially at scales larger than $L$.

We resume the results on the flow around small particle in the Bousinessq flow with $L\ll l_d$ but $L\gtrsim l_d(1-\gamma_z/\gamma)^{1/4}$. The flow is very close to stratified flow with $\bm \gamma=\gamma{\hat z}$. The perturbation flow around the particle decays (quasi) exponentially at scales larger than $L$ where the streamlines are closed. The usual Stokes flow with constant correction holds at scales much smaller than $L$.

We observe that the range of validity of the perturbation flow discovered in \cite{Ardekani} is very narrow. The derivation of the flow presumes that the gradient $\bm \gamma$ is constant over the scale $L$ so that
$L\ll l_d$. The stability condition $L\gtrsim l_d(1-\gamma_z/\gamma)^{1/4}$ implies then that $(1-\gamma_z/\gamma)^{1/4}\ll 1$. This implies $(1-\gamma_z/\gamma)\lesssim 10^{-4}$ meaning very small deviations from purely vertical stratification, the situation that seems unlikely to occur in typical marine environment. Thus, we pass to consider the typical situations where the flow is turbulent with $(1-\gamma_z/\gamma)^{1/4}\sim 1$. \\

\section{\label{flow}Scalar advection-diffusion equation for flow around small swimmer in turbulence}

In this Section we study the flow around small swimmer translating in the turbulent flow. We demonstrate that in the limit of large Prandtl numbers, which occurs in practical applications often, the problem can be reduced to scalar advection-diffusion equation.

The limit $\mathrm{Pr}\gtrsim 1$ occurs in natural environments often. The relations discussed in the Section on relations in turbulence imply that in this case $L\sim l_d \mathrm{Nu}^{1/8}$. One finds $L\sim l_d$ unless the Nusselt number is unrealistically large. In the turbulent flow with $\mathrm{Pr}\gtrsim 1$ the following hierarchy of scales holds $l_{\nu}\gtrsim l_d\sim L$.

Specifying to $\mathrm{Pr}^{1/2}\gg 1$ where $l_{\nu}\gg l_d$ one can use Eqs.~(\ref{start}). Further since the characteristic time of variations of $\bm w'$ is $\lambda^{-1}$ then we can neglect $\partial_t\bm  w'$ term in the first of Eqs.~(\ref{start}) in comparison with the viscous term. We obtain
\begin{eqnarray}&&\!\!\!\!\!\!\!\!\!
0=-\nabla P+\Theta \bm g+\nu\nabla^2\bm w'+f{\hat z}\delta(\bm x),\label{start0}\\&&\!\!\!\!\!\!\!\!\!
\partial_t\Theta+\sigma\bm x\cdot \nabla\Theta+\bm w'\cdot\nabla\theta_0\left(\bm x+\bm y\left[t\right]\right)=\kappa\nabla^2\Theta. \label{start00}
\end{eqnarray}
We stress that this system provides valid description to the flow around small particle in arbitrary Bousinessq turbulence with large Prandtl number.
Turbulent transport is described by the material derivative term $\partial_t\Theta+(\sigma\bm r\cdot\nabla)\Theta$ that occurs universally in the description of the advection of the passive scalar fields by turbulence at large Prandtl numbers, see \cite{review,Batchelor} and references therein.
This term is comparable with the diffusive one at a characteristic scale $\ell_d=\sqrt{\kappa/\lambda}$ and it is dominating at larger scales.

We note that Eq.~(\ref{start0}) coincides with Eq.~(\ref{s1}). Consequently the solution is given by Eq.~(\ref{a15}). We find that the flow is described by the scalar advection-diffusion equation
\begin{eqnarray}&&\!\!\!\!\!\!\!\!\!
\partial_t\Theta'+\sigma\bm x\cdot \nabla\Theta'=\kappa\nabla^2\Theta'+\nabla_i\theta_0\left(\bm x+\bm y\left[t\right]\right)
\label{form}\\&&\!\!\!\!\!\!\!\!\!
 \int \frac{d\bm k}{(2\pi)^3}\frac{g\Theta'(\bm k) \left[\delta_{iz}-{\hat k}_i{\hat k}_z\right]}{\nu k^2}\exp\left[-i\bm k\cdot\bm x\right]+\frac{\kappa f\nabla^2\delta(\bm x)}{g}, \nonumber
\end{eqnarray}
where we used that the Fourier space relation $\Theta'\equiv\Theta-f/g$ implies $\Theta'\equiv\Theta-f\delta(\bm x)/g$ in the real space ($f$ is considered time-independent). We used $\sigma\bm x\cdot \nabla \delta(\bm x)=\nabla\cdot\left[\sigma\bm x \delta(\bm x)\right]=0$. Once $\Theta'$ is found the pressure and velocity can be obtained by the Fourier space relations
\begin{eqnarray}&&\!\!\!\!\!\!\!\!\!\!\!\!\!\!\!
P(\bm k)=\frac{i g k_z\Theta'(\bm k)}{k^2},\ \ w'_i(\bm k)= -\frac{g\Theta'(\bm k) \left[\delta_{iz}-{\hat k}_i{\hat k}_z\right]}{\nu k^2}.\label{form10}
\end{eqnarray}
We conclude that the problem of finding the flow around small particle in the Boussinesq flow with large $\mathrm{Pr}$ is described by one integro-differential equation of advection-diffusion type with point-like source. Clearly the difficulty in solving this equation is in the spatial dependence of $\nabla_i\theta_0\left(\bm x+\bm y\left[t\right]\right)$. Progress in the solution demands simplifications of the latter term.

\section{Correction to the Stokeslet flow at $x\ll L$}

We consider the flow at scales $x\ll L$ where it has to be close to the usual Stokeslet flow. Our purpose in this Section is to find the correction to that flow.

We observe that due to $x\ll l_d$ one can use $\nabla\theta_0\left(\bm x+\bm y\left[t\right]\right)\approx -\bm \gamma(t)$ with $\bm \gamma=-\nabla\theta_0\left(\bm y\left[t\right]\right)$. Using the incompressibility $tr\: \sigma=0$ we find in Fourier space
\begin{eqnarray}&&\!\!\!\!\!\!\!\!
\partial_t\Theta'-(\sigma^t\bm k\cdot\nabla)\Theta'=-\alpha(\bm k)\Theta'-\kappa k^2 f/g,\nonumber\\&& \!\!\!\!\!\!\!\!
\alpha(\bm k)\equiv \kappa k^2d(\bm k),\ \ d(\bm k)\equiv
1+\frac{g{\tilde \gamma}}{\kappa \nu k^4}.\label{dif}
\end{eqnarray}
To find the solution we pass to the moving frame
$\tilde {\Theta}(\bm k, t)=\Theta'(\bm k(t), t)$ where $\bm k(t)\equiv W^{-1, t}(t)\bm k$ with
\begin{eqnarray}&&\!\!\!\!\!\!\!\!\!\!
\!\!\!\!\!\!\!\!\!\!\dot{W}=\sigma W,\ \ \dot{W}^{-1, t}=-\sigma^t W^{-1, t},\ \ W_{ij}(t\!=\!0)\!=\!
\delta_{ij}
. \label{evol}
\end{eqnarray}
The matrix $\sigma$, and thus $W$, have to be considered random for turbulence and described statistically. The properties of the statistics of $W$ that are relevant here do not depend on the details of the statistics of $\sigma$ due to universality \cite{review}, yet for clarity we assume that the statistics of $\sigma$ is close to the Lagrangian statistics of $\nabla_j u_i$ (the statistics in the frame of fluid particle). This holds provided the velocity of the particle's drift with respect to the flow is much smaller than the characteristic velocity $u_{\eta}\sim \lambda\ell_{\eta}$ of the viscous scale eddies of turbulence \cite{Frisch}. It seems that this assumption is not restrictive and it is obeyed in typical natural situations.
Since $\sigma$ is statistically the same as the velocity gradient of $\bm u$ in the fluid particle's frame, then $W$ is statistically the same as the Jacobi matrix of the turbulent flow backward in time \cite{review}. That is, if we consider the Lagrangian trajectories $\bm q(t, \bm r)$ defined by
$\partial_t \bm q(t, \bm r)=\bm u[t, \bm q(t, \bm r)]$ and $ \bm q(t=0, \bm r)=\bm r$,
then $W_{ij}(t, \bm r)=\partial_j q_i(t, \bm r)$ at $t<0$ describes the evolution of small volumes in the turbulent flow backward in time and obeys Eq.~(\ref{evol}).
In particular, since the Lyapunov exponents of the backward in time flow are $(-\lambda_3, -\lambda_2, -\lambda_1)$ where $(\lambda_1, \lambda_2, \lambda_3)$ are the Lyapunov exponents of the forward in time flow, then $k(t)$, which is governed by $W^{-1, t}$ rather than $W(t)$, obeys
\begin{eqnarray}&&
\lim_{t\to -\infty}(1/|t|) \ln [k(t)/k(0)]=\lambda_1, \label{lim}
\end{eqnarray}
see details in \cite{review}. Thus the growth of $\bm k(t)$ with $|t|$ is similar to the exponential growth of the separation between two infinitesimally close fluid particles in turbulence (governed by the principal Lyapunov exponent $\lambda_1$).

The limit in Eq.~(\ref{lim}) holds for almost every realization of $\sigma(t)$ and does not involve the randomness of turbulence that disappears after taking the infinite time limit. To describe the fluctuations of $\bm k(t)$ when $t$ is finite, one introduces the polar representation  $\bm k(t)=k\exp[\rho(t)]{\hat n}(t)$, where $|{\hat n}|=1$. Using
$\dot {\bm k}=-\sigma^t \bm k$ one finds \cite{review}
\begin{eqnarray}&&
\dot {\hat n}=-\sigma^t {\hat n}+{\hat n}\zeta,\ \ {\dot \rho}=\zeta,\ \ \zeta\equiv -{\hat n}\sigma{\hat n}.
\end{eqnarray}
It follows that $\ln[k(t)/k]=\int_t^0 \zeta(t') dt'$ where $\zeta$ is a finite-correlated noise which correlation time $\tau_c$ is of order of the correlation time of $\sigma$, so that $\tau_c\sim \lambda^{-1}$. Thus Eq.~(\ref{lim}) resembles the law of large numbers. To find the moments of $k(t)$ one introduces
\begin{eqnarray}&&
\lim_{t\to-\infty}(1/|t|) \ln \langle k^l(t)\rangle\equiv \varphi(l). \label{gamma}
\end{eqnarray}
The function $\varphi(l)$ is convex and obeys $\varphi(0)=\varphi(-3)=0$, so it is negative at $-3<n<0$ and positive otherwise. This holds independently of the statistics of turbulence (see \cite{review} for details). In the moving frame Eq.~(\ref{dif}) becomes
\begin{eqnarray}&&
\!\!\!\!\!\!\!\!\partial_t \tilde {\Theta}=-\alpha[\bm k(t)]\tilde {\Theta}-\kappa k^2(t) f/g.
\end{eqnarray}
We consider $\tilde {\Theta}$ at $t=0$, taking the initial condition at $t=-T$ and studying the limit $T\to\infty$, i. e. we focus on the steady state solution. Using $\tilde {\Theta}(t=0)=\Theta'(t=0)$,
\begin{eqnarray}&&\!\!\!\!\!\!\!\!\!\!
\!\!\!\!\!
\Theta'\!=-\frac{\kappa f}{g}
\!\!\int_{-\infty}^0 \!dt \exp\left[-\!\!\int_t^0\alpha\left[\bm k(t')\right]dt'\right]k^2(t)
.\label{a41}
\end{eqnarray}
This together with Eq.~(\ref{form10}) is implicit Fourier space solution to the problem of finding the flow around small particle at small distances.

The exponent $-\alpha\left(\bm k\right)$ has to coincide in the limit with the eigenvalues $\lambda(\bm k)$ of the linearized problem. This is indeed so: considering
Eq.~(\ref{labd}) in the limit of large Prandtl numbers $\nu\gg \kappa$ one has to leading order
\begin{eqnarray}&&
\lambda(\bm k)\approx\frac{\nu k^2}{2}\left[\sqrt{1-\frac{2\kappa}{\nu}-\frac{4g {\tilde \gamma}}{\nu^2k^4}}-1\right]-\frac{\kappa k^2}{2}
\nonumber\\&&
\approx -\kappa k^2 -\frac{g {\tilde \gamma}}{\nu k^2}=-\alpha\left(\bm k\right),
\end{eqnarray}
where we used $g {\tilde \gamma}/[\nu^2k^4]\sim [\mathrm{Pr} L^2 k^2]^{-2}\ll 1$.

The solution (\ref{a41}) can be rewritten using $\bm k(t)=k\exp[\rho(t)]{\hat n}(t)$
\begin{eqnarray}&&
\!\!\!\!\!\!\!\!
\Theta'\!=-\frac{\kappa k^2 f}{g}
\!\!\int_{-\infty}^0 \!dt \exp\left[s(t)\right],\ \
s(t)=2\rho(t)\label{res}\\&&\!\!\!\!\!\!\!\!
-\kappa k^2 \int_t^0  \exp\left[2\rho(t')\right] dt'
-\frac{g}{\nu k^2}\int_t^0 {\tilde \gamma}(t')\exp\left[-2\rho(t')\right]
dt',\nonumber
\end{eqnarray}
where ${\tilde \gamma}(t)=\gamma_z(t')-n_z(t')\left[\bm \gamma(t')\cdot{\hat n}(t')\right]$ can be considered stationary noise with amplitude of the order of the characteristic amplitude of the fluctuations of $\gamma$.
The characteristic time of variations of $\bm \gamma(t)$ is the characteristic time-scale $\lambda^{-1}$ of variations of $\gamma(t, \bm x)$ in the fluid particle's frame. Inertia is considered small here so that the deviation of the particle's trajectory from the one of the fluid particle is negligible.

To understand the behavior of $s(t)$ qualitatively we use the mean field approximation $\rho(t)=\lambda_1|t|$
\begin{eqnarray}&&
s(t)\approx-\kappa k^2 \int_{t}^0  dt' \left[e^{2\lambda_1|t'|}
+\frac{g{\tilde \gamma}(t')\exp\left(-2\lambda_1|t|\right)}{\kappa \nu k^4}\right]
\nonumber\\&&+2\lambda_1|t|. \label{s}
\end{eqnarray}
When $\lambda_1 |t|\ll 1$ this reduces to
\begin{eqnarray}&&\!\!\!\!\!\!\!\!\!\!\!\!\!\!
s(t)\!\approx\! 2\lambda_1|t|\!-\!\kappa k^2 |t|\!-\!\frac{g {\tilde \gamma}(0)|t|}{\nu k^2}\!=\![\lambda(\bm k, 0)+2\lambda_1]|t|,\label{integral}
\end{eqnarray}
where we introduced
\begin{eqnarray}&&
\lambda(\bm k, t)=-\kappa k^2 -\frac{g {\tilde \gamma}(t)}{\nu k^2}.
\end{eqnarray}
Since we consider scales $l_d\gg x$ where $\kappa k^2\gg g {\tilde \gamma}/ \nu k^2$ and $\kappa k^2\gg \lambda_1$ then $-\kappa k^2|t|$ term dominates $s(t)$. In particular $s(t)$ is decreasing function of $t$ which corresponds to $\lambda(\bm k)<0$ at $k\gg k_0$, see the Section on stability.
Thus the integral in Eq.~(\ref{res}) converges at the time-scale $[\kappa k^2]^{-1}$ which is much smaller than $\lambda^{-1}$ by $k^2l_d^2\gg 1$. The leading order formula for $\Theta'$ is found by setting $s(t)\approx \dot s(0) t$ in the integral in Eq.~(\ref{res}). One finds $\Theta'\approx -\kappa k^2 f/[g \dot s(0)]$ which gives
\begin{eqnarray}&&
\!\!\!\!\!\!\!\!
\Theta'(\bm k)\!\approx -\frac{f L^4(0)k^4}{g \left[L^4(0)k^4+1-2L^4(0)\bm k\sigma(0)\bm k/\kappa\right]},\label{solution}
\end{eqnarray}
where we used $\zeta(0)=-{\hat k}\sigma{\hat k}$ and defined $L^4(t)=\kappa \nu/[g {\tilde \gamma}(t)]$. This differs from solution (\ref{b12}) in the stably stratified fluid by the use of the instantaneous value $\gamma(t=0)$ in the definition of ${\tilde \gamma}$ and by the presence of the transport term involving the turbulent velocity gradient at the position of the particle $\sigma(0)$.

The solution (\ref{solution}) describes well the flow at $k^4L^4\gg 1$ where the first term dominates the denominator,
\begin{eqnarray}&&
\!\!\!\!\!\!\!\!
\Theta'(\bm k)\!\approx -\frac{f}{g},\ \ k^4L^4\gg 1.
\end{eqnarray}
It is readily seen that this solution produces the usual Stokeslet flow that has to hold at $x\ll L$. The leading order correction to the Stokeslet flow is given by
\begin{eqnarray}&&
\!\!\!\!\!\!\!\!
\Theta'(\bm k)\!\approx -\frac{f }{g} +\frac{f }{gL^{4}(0)k^{4}} -\frac{2\bm k\sigma(0)\bm k f }{g\kappa k^4},\label{solution1}
\end{eqnarray}
where the second term is kept to demonstrate how the variation of $\bm \gamma$ in comparison with the vertical direction is present in the correction. In fact the ratio of the last term to the second one is $~ L^4 k^2 l_d^{-2}\gg 1$ so that the consistent form of $\Theta'(\bm k)$ to the lowest order in $kL\gg 1$ is
\begin{eqnarray}&&
\!\!\!\!\!\!\!\!
\Theta'(\bm k)\!\approx -\frac{f }{g}-\frac{2\bm k\sigma(0)\bm k f }{g\kappa k^4},\ \ kL\gg 1. \label{correction1}
\end{eqnarray}
Thus it is the turbulent transport that dominates the correction to the Stokeslet flow at small scales and not the scalar gradients $\bm \gamma$.

The correction is a random field that depends on the random value of the matrix of velocity gradients $\sigma$ at the location of the particle. One has  $\langle \sigma_{ij}\rangle=
\langle \nabla_j u_i\rangle=0$ and $\langle \sigma_{ij}\sigma_{mn}\rangle=\langle \nabla_ju_i\nabla_n u_m\rangle$, where the last average can be taken in the Eulerian frame due to incompressibility (we use here that the statistics of $\sigma$ is assumed to be close to the one of the matrix of velocity gradients in the frame of the fluid particle).
One finds
\begin{eqnarray}&&\!\!\!\!\!\!\!\!
30\nu \langle \sigma_{ij}\sigma_{mn}\rangle=\epsilon\left[4\delta_{im}\delta_{jn}-\delta_{in}\delta_{jm}-\delta_{ij}\delta_{mn}\right].
\end{eqnarray}
where isotropy (typically valid for small-scale turbulent fluctuations that determine $\nabla \bm u$ \cite{Frisch}) and spatial uniformity are assumed.
The form of $\langle \nabla_ju_i\nabla_n u_m\rangle$ is fixed uniquely by the demands of isotropy, incompressibility and spatial uniformity that imply
 $\langle \nabla_m u_i\nabla_n u_m\rangle=\langle \nabla_m \left[u_i\nabla_n u_m\right]\rangle=0$. The relation is {\it exact} due to stationarity condition
 $\nu \langle \nabla_ju_i\nabla_j u_i\rangle=\epsilon$. Thus we obtain the {\it exact} result for fluctuations $\delta\Theta=\Theta'+f/g$ of $\Theta'$ near
 the Stokeslet value $-f/g$ (clearly $\langle \delta\Theta\rangle=0$) due to background turbulence,
 \begin{eqnarray}&&\!\!\!\!\!\!\!\!\!\!
\langle \delta\Theta^2\rangle^{1/2}\!=\frac{2f}{g\kappa k^2}\sqrt{\frac{\epsilon}{15\nu}}.\label{fluct}
\end{eqnarray}
For the r. m. s. deviation near the average Stokeslet value we have
 \begin{eqnarray}&&\!\!\!\!\!\!\!\!\!\!
\frac{\langle \delta\Theta^2\rangle^{1/2}}{\langle \Theta'\rangle}=\frac{2}{\kappa k^2}\sqrt{\frac{\epsilon}{15\nu}}\sim\frac{1}{l_d^2k^2},
\end{eqnarray}
where we used that $l_d^2\sim \kappa/\sqrt{\epsilon/\nu}$ when $\mathrm{Pr}\gtrsim 1$. Clearly the condition of validity of the derivation $kL\gg 1$ implies the correction is small due to $L\sim l_d$. Note however that
using Eq.~(\ref{correction1}) one can discuss the impact of strong fluctuations of $\nabla\bm u$ above the typical value $\sqrt{\epsilon/\nu}$ that occur in the turbulent flow quite often due to intermittency.

We stress that Eqs.~(\ref{correction1}),(\ref{fluct}) provide the exact leading order corrections at scales $x\ll L$ to the Stokeslet flow around the particle translating in arbitrary (turbulent) Bousinessq flow.

\section{Strong deviations from Stokeslet: flow at scales $L\sim l_d$}

Flow around small particle presents the highest interest at scales $x\sim L\sim l_d$ where the deviations from the Stokeslet flow are not small. Particularly, one would like to know if the features observed in the flow in the fluid at rest, i.e., fast decay at scales larger than $L$ and toroidal eddies, would hold in turbulence too. To address this question one needs to find the complete solution to Eq.~(\ref{form}), which is formidable. Therefore, we resort to order-of-magnitude calculation.

We observe that the flow obtained by disregarding the spatial variation of $\nabla\theta_0\left(\bm x+\bm y\left[t\right]\right)$ and using $\nabla\theta_0\left(\bm x+\bm y\left[t\right]\right)\approx -\bm \gamma(t)$ is to reproduce the qualitative features of the flow well. Thus we study the solution (\ref{res}) at $x\sim L$. In this case the three terms in Eq.~(\ref{integral}) are of the same order due to $kl_d\sim kL \sim 1$. We observe that the integral for $\Theta'$ in Eq.~(\ref{res}) converges over time-scales of order $\lambda_1^{-1}$ because at larger times $s(t)\sim -k^2l_d^2\exp[2\lambda_1|t|]$ where $\kappa/\lambda_1\sim l_d^2$. Since ${\tilde \gamma}(t)$ varies over time-scale $\lambda_1^{-1}$ then by order of magnitude we can put ${\tilde \gamma}(t)={\tilde \gamma}(0)$ in  Eq.~(\ref{s}) which gives
\begin{eqnarray}&&\!\!\!\!\!\!\!\!\!\!
s(t)\approx-\kappa k^2 \int_{t}^0  dt' \left[e^{2\lambda_1|t'|}
+\frac{g{\tilde \gamma}(0)e^{-2\lambda_1|t|}}{\kappa \nu k^4}\right]+2\lambda_1|t|
\nonumber\\&&\!\!\!\!\!\!\!\!\!\!\approx\! 2\lambda_1|t|\!-\!\frac{\kappa k^2}{2\lambda_1} \left[e^{2\lambda_1|t|}\!-\!1\right]
-\frac{g{\tilde \gamma}(0)\left[1\!-\!e^{-2\lambda_1|t|}\right]}{2 \lambda_1\nu k^2}
.
\end{eqnarray}
We find
\begin{eqnarray}&&
\!\!\!\!\!\!\!\!
\Theta'=-\frac{\kappa k^2 f}{2g\lambda_1}
\!\!\int_0^{\infty} \!dt \exp\left[t\!-\!\frac{\kappa k^2}{2\lambda_1} \left[e^{t}\!-\!1\right]
\!-\!\frac{g{\tilde \gamma}(0)\left[1\!-\!e^{-t}\right]}{2 \lambda_1\nu k^2}\right].\nonumber
\end{eqnarray}
Introducing the variable $y=\exp[t]$ we find
\begin{eqnarray}&&
\!\!\!\!\!\!\!\!
\Theta'\!=\!-\!\frac{\kappa k^2 f\exp\left[p-r\right]}{2g\lambda_1}
\!\!\int_1^{\infty} \!dy \exp\left[-p y+\frac{r}{y}\right].
\end{eqnarray}
where we introduced $p=\kappa k^2/[2\lambda_1]$, $r=g{\tilde \gamma}(0)/[2 \lambda_1\nu k^2]$.
The corresponding result for the velocity is, see Eq.~(\ref{form10}),
\begin{eqnarray}&&\!\!\!\!\!\!\!\!\!\!\!\!\!\!
w'_i\!=\! \frac{\kappa  f\exp\left[p\!-\!r\right]\left[\delta_{iz}\!-\!{\hat k}_i{\hat k}_z\right]}{2\nu\lambda_1}\!\!\int_1^{\infty} \!dy \exp\left[\!-p y\!+\frac{r}{y}\right].\label{final}
\end{eqnarray}
Thus, for the real-space velocity profile we find
\begin{eqnarray}&&\!\!\!\!\!\!\!\!
w'_i({\tilde l}_d\bm x)\!=\! \frac{f}{\nu{\tilde l}_d}\!\!\int_1^{\infty}\! \!dy\int_{q>0.3}\frac{d\bm q}{(2\pi)^3} \left[\delta_{iz}\!-\!{\hat q}_i{\hat q}_z\right]\exp\left[i\bm q\cdot\bm x\right]\nonumber\\&&\!\!\!\!\!\!\!\!
\exp\left[\!-q^2 (y-1)\!-\frac{{\tilde l}_d^4\left[{\hat \gamma}_z-({\hat \gamma}\cdot {\hat q}){\hat q}_z\right] (y-1)}{L^4 q^2 y}\right],\label{final1}
\end{eqnarray}
where in the inverse Fourier transform we passed to the integration variable $\bm q=\bm k{\tilde l}_d$ with ${\tilde l}_d=\sqrt{\kappa/[2\lambda_1]}$ of the order of $l_d$. We introduced $L^4(0)=\kappa \nu/[g\gamma(0)]$ and $\bm \gamma=\gamma{\hat \gamma}$. The (fluctuating) factor $[{\tilde l}_d/L(0)]^4$ is of order one. The direction ${\hat \gamma}$ is statistically isotropic because the small-scale statistics of turbulence is. We disregard that $\bm \gamma$ has non-vanishing average because that is much smaller than the typical value of $\bm \gamma$.

The formula (\ref{final}) is valid for $kL\gtrsim 1$. Hence it can be used to find the qualitative picture of the flow (in real space) at scales $x\lesssim L$. Thus, in the inverse Fourier transform one has to put a cutoff at small $kL$ where the formulas become invalid. Since $L\sim {\tilde l}_d$ then this cutoff corresponds to $q\gtrsim 1$ in Eq.~(\ref{final1}). This cutoff is necessary because ${\tilde \gamma}$ can be negative producing exponential divergence of the integral at small $q$. It follows that for the directions in Fourier space for which ${\tilde \gamma}$ is negative the integral is determined by the cutoff.

It is clear that the obtained solution involves order one change of the solution corresponding to the fluid at rest. Thus with finite probability there will be closed streamlines in the flow. The streamlines around the swimmer will fluctuate occasionally getting open. Though in the absence of degeneracy forbidding closed streamlines our conclusions seems highly plausible, further studies are needed.

\section{Flow around swimmers}

The fundamental Stokeslet solution described by Eqs.~(\ref{start0})-(\ref{start00}) corresponds to a flow induced by a small particle moving under the action of an external force (e.g. passive sinking under the action of gravity). The flow field around self-propelled objects could be, however, quite different from that of a passively towed particle. A steadily self-propelled swimmer generates no net momentum flux, since the thrust is counter-balanced by the drag force (in contrast to the particle which steady motion demands an external force, the swimmer can move on its own). The two forces of equal magnitude, acting in opposite directions and separated by some distance, constitute a force dipole of strength ${\hat s}$, so that the flow induced by a self-propelled swimmer can be approximately described by
\begin{eqnarray}&&
\nabla P_s=\Theta_s \bm g +\nu\nabla^2\bm w_s'+{\hat s}_{ij}\nabla_j \delta(\bm r),
\end{eqnarray}
where  ${\hat s}_{ij}$ is a diagonal matrix ${\hat s}_{ij}=diag[-{\hat s}/3, -{\hat s}/3, 2{\hat s}/3]$, and the subscript ``s" will denote the fields of the force-doublet flow. Passing to
${\tilde P}_s\equiv P_s+{\hat s}\delta(\bm r)/3$ we find that the flow around the swimmer is determined from
\begin{eqnarray}&&
\nabla {\tilde P}_s=\Theta_s \bm g +\nu\nabla^2\bm w_s'+{\hat s}\frac{\partial \delta(\bm r)}{\partial z}{\hat z},\nonumber\\&&
\partial_t\Theta_s+(\sigma\bm r\cdot\nabla)\Theta_s-\bm \gamma \cdot \bm w'_{{\hat s}z}=\kappa\nabla^2\Theta_s, \label{s10}
\end{eqnarray}
The relation between the Stokeslet and the force-doublet flows is trivial without turbulence when the term $\partial_t\Theta+(\sigma\bm r\cdot\nabla)\Theta$ is missing in the equations. The equations corresponding to the flow induced by the force-doublet can be simply obtained by differentiating the equations for Stokeslet with respect to $z$, so $\bm w'_s=({\hat s}/f)\partial_z \bm w$, $\Theta_s=({\hat s}/f)\partial_z \Theta$ and ${\tilde P}_s=({\hat s}/f)\partial_zP$. When turbulence is considered, the coefficients in Eqs.~(\ref{s10}) depend on the coordinate, so differentiation over $z$ does not produce Eqs.~(\ref{s1}). Thus in the presence of turbulence, the relation between the Stokeslet and the force-doublet flows is non-trivial. Nevertheless qualitatively the conclusions reached for the Stokeslet continue to hold for the doublet. This can be demonstrated using the corresponding expressions for pressure and velocity are (we write directly the pressure $P_s$, rather than ${\tilde P}_s$),
\begin{eqnarray}&&\!\!\!\!\!\!\!\!\!\!\!\!\!\!\!
P_s\!=\!\frac{ig k_z\Theta_s+k_i {\hat s}_{ij}k_j}{k^2},\ \ \nu k^2\bm w_s'\!=\!\Pi(\bm k)\left[\Theta_s \bm g+i {\hat s} \bm k\right].\label{swim}
\end{eqnarray}
The calculations repeat the steps used for the Stokeslet.

\section{Conclusion}

In this work we derived the stability condition that holds for arbitrary Boussinesq flow which is smooth below a certain scale $l_d$. This is generalization of the classical convective stability condition which derivation is possible thanks to universal spatial behaviour of the flow at scales smaller than $l_d$. The flow there can be described by linear profile corresponding to the first term in the Taylor series.

The condition is the demand that small perturbations below the scale $l_d$ decay in order to preserve the laminarity of the flow in that region (at larger scales the flow can be turbulent). This involves the Rayleigh scale $L=(\nu\kappa/g\gamma)^{1/4}$ and the dimensionless measure of fluctuations of the gradient $\bm \gamma$ of the scalar around the vertical direction, $\mathrm{Fl}=(1-\gamma/\gamma_z) ^{1/4}$. We prove that the flow with $L/l_d\ll \mathrm{Fl}$ is unstable with respect to growth of small fluctuations at scales smaller than $l_d$.

Thus there is a non-trivial condition $L/l_d\gtrsim \mathrm{Fl}$ on arbitrary Boussinesq flow whenever the gradient $\bm \gamma$ deviates from strict stratification condition $\mathrm{Fl}=0$. Since deviations from strict verticality are inevitable in nature then we conclude that this condition needs to be verified all along when dealing with natural environments.

The first conclusion that we draw from the stability condition is that quiescent stratified fluids are unstable in the limit of large Rayleigh numbers, $\mathrm{Ra}$. This is because $\mathrm{Fl}$ is finite in nature.

This poses the question of finding the realistic behaviour of stratified fluid in $\mathrm{Ra}\to\infty$ limit. To address this question we propose introduction of the random source of fluctuations of $\bm \gamma$  into the equations and look for a new steady state. Our work implies that this state is likely not to be close to the state of quiescent stratified fluid (unless the non-linear terms arrest the fluctuations' growth at weak non-linearity which seems unlikely). This new steady state may shed light on the puzzle of the layered structure of the density in the ocean which origin is not yet fully understood \cite{thorpe05}. It is left for future work.

Applying the $L/l_d\gtrsim \mathrm{Fl}$ condition to the study of the turbulent Boussinesq flow we find that the previously derived phenomenological relations \cite{Lohse} can be understood as the condition of stability of the flow when the Prandtl number, $\mathrm{Pr}$, is not small and the Nusselt number $\mathrm{Nu}$ is not too large. Similarly to how in the Navier-Stokes turbulence one can understand the viscous scale as the one at which the flow becomes laminar (the Reynolds number at the viscous scale is of order one \cite{Frisch}) so in this case of Boussinesq turbulence the dissipation scale $l_d$ can be understood as the scale at which the dissipative processes regularize the convective instability. In contrast, the phenomenological relations diverge from the stability condition in the limits $\mathrm{Nu}^{1/8}\gg 1$ or $\mathrm{Pr}\ll 1$. The reason for this divergence is to be studied in the future.

Finally we consider the implications of the stability condition to the flow of small particle in arbitrary background flow. The small particle can be considered as the source of the flow fluctuations. The stability with respect to those fluctuations implies that the recently found solution for the flow around small particle in quiescent stratified fluid \cite{Ardekani} seems to be highly limited in scope.

It was demonstrated that the flow around small particle moving in the quiescent stratified fluid with constant gradient $\bm \gamma=\gamma{\hat z}$ has a non-trivial structure of closed streamlines at the scale $L$ beyond which the flow decays rapidly \cite{Ardekani}. However, the constancy of the gradient implies the condition $L\ll l_d$ of observability of this flow. Using the stability condition derived in this work we conclude that the flow is observable if $l_d \mathrm{Fl}\lesssim L\ll l_d$ which implies very small deviations of the gradients from verticality, $\mathrm{Fl}\ll 1$.

Considering the flow with $l_d \mathrm{Fl}\lesssim L\ll l_d$ we describe quantitatively the rate of decay of the flow at scales larger than $L$ (this decay was addressed in \cite{Ardekani} numerically without quantitative conclusions). We demonstrate that the flow decays inversely with the distance to the power nine. Due to the high exponent of the power law this algebraic decay seems to be practically indistinguishable from the exponential one, which we prove by computing the solution numerically.

The regime with $\mathrm{Fl}\ll 1$ is likely to hold very rarely, if at all, in typical marine environments. This is due to intrinsic presence of fluctuations in stratification gradient bringing finite $\mathrm{Fl}$. This poses the question of finding the flow around small particle in Boussinesq flows which is relevant to nature.

We consider the most practical case of the particle translating in the turbulent Boussinesq flow. There are two regions in the perturbation flow caused by the particle. Below the Rayleigh scale the flow is close to the usual Stokes flow. We succeeded in deriving analytic formulas for the correction to the Stokes flow in that region demonstrating that the leading order correction is a small constant drift. Since drift correction can accumulate bringing finite effects this poses the question of practical implications of this correction. Direct generalization of our calculation can be used to describe the corrections to the Stokes force on the particle due to stratification.

In contrast at the Rayleigh scale the flow is very different from the Stokes flow. We demonstrated that in the practically relevant regime of large Prandtl numbers the flow can be described by one scalar integro-differential advection-diffusion equation (\ref{form}) instead of the complete system of four hydrodynamic equations. We propose this equation as efficient tool for the numerical study of the flow round small particles in stratified turbulent flow.

Though the closed-form solution to that equation cannot be obtained, we succeed in constructing the solution that holds by the order of magnitude providing valid qualitative description of the solutions. This indicates that the closed streamlines holding for quiescent stratified background persist for the turbulent background with finite probability. The structure of those lines is different being random though. The numerical study of the statistics of the flow near small particles in Boussinesq turbulence and the corresponding probabilities is left for future work.

We conclude by providing typical numbers for the flow around small swimmer in various aquatic environments. Using the extreme value of the density gradient $\gamma\rho_0=1$ kg m$^{-4}$ \cite{Ardekani} that may occur locally in fjords \cite{alldredge02}, lakes and reservoirs \cite{patterson84} with $\mu=10^{-3}$ kg~m$^{-1}$~s$^{-1}$ yields $L\approx 0.6$~mm for salt-stratified water ($\kappa\approx 1.3\times10^{-9}$~m$^2$ s$^{-1}$) and $L\approx 2$~mm for temperature-stratified water ($\kappa\approx 1.4\times10^{-7}$~m$^2$ s$^{-1}$). Further considering weakly turbulent conditions with the dissipation rate per unit mass $\epsilon \approx 10^{-10}$~m$^2$ s$^{-3}$ (e.g. Kunze et al.\cite{kunze06} measured $\epsilon \lesssim 10^{-9}$ in a coastal inlet) gives $\lambda=\sqrt{\epsilon/\nu} \approx 0.01$~s$^{-1}$. Thus, the corresponding values of $L/l_{dif}$ are $\approx 0.5$ and $\approx 1.7$ for temperature- and salt-stratified water, respectively. Furthermore, in the marine environment the buoyancy frequency $N=\sqrt{g\gamma}$ corresponding to the marginal oscillations which the stable stratification supports \cite{thorpe05} is typically in the range between $10^{-4}$ and $10^{-2}$~s$^{-1}$, yielding density gradients $\gamma\rho_0$ that ranges between $10^{-6}$ and $10^{-2}$~kg m$^{-4}$, several orders of magnitude lower than that considered in \cite{Ardekani}. In some extreme cases, however (e.g. during seasonal thermocline \cite{thorpe05}) $N$ may exceed $0.05$~s$^{-1}$ so $\gamma\rho_0$ may reach $\approx 0.3$~kg m$^{-4}$. Using this extreme value of density stratification and $\epsilon \approx 10^{-10}$~m$^2$ s$^{-3}$ we arrive at $L/l_{dif} \approx 0.7$ and $L/l_{dif}\approx 2.3$ for temperature- and salt-stratified water, respectively. For the less extreme conditions of marine turbulence and/or stratification typically $L/l_{dif}>1$. For example, for $\epsilon \approx 10^{-9}$~m$^2$ s$^{-3}$ and $\gamma\rho_0\approx 0.01$~kg m$^{-4}$ we find $L/l_{dif} \approx 2.8$ and $L/l_{dif} \approx 9.5$ for temperature and salt stratification, respectively. These estimates demonstrate clearly that the relevant physical situation is the one with $L\gtrsim l_d$ where the solutions with turbulence are to be considered.

We thank the anonymous Referee for very useful comments that led us to revisit the properties of the Boussinesq flow. We would like to thank John Dabiri (Caltech) for for stimulating discussions on the subject.
This work was supported by the US-Israel Binational Science Foundation (BSF) via the Transformative Science Grant \#2011553 and by Japan Technion Society Research Fund.

\end{document}